\documentclass[preprints,article,accept,moreauthors,pdftex]{Definitions/mdpi}
\usepackage{booktabs}
\usepackage{multirow}
\usepackage{soul}
\usepackage{microtype}
\graphicspath{{Definitions/}}
\usepackage{upgreek}

\newcommand{\apj}{Astrophys. J.}

\newcommand{\apjl}{Astrophys. J. Lett.}

\newcommand{\aj}{Astron. J.}

\firstpage{1}
\pubvolume{xx}
\issuenum{1}
\articlenumber{5}
\pubyear{2019}
\copyrightyear{2019}
\history{Received: 30 June 2019; Accepted: 4 December 2019; Published: date}
\updates{yes} 

\Title{High Energy Radiation from Spider Pulsars}

\Author{Chung Yue Hui $^{1,*}$\orcidA{} and Kwan Lok Li $^{2}$ }
\AuthorNames{Chung Yue Hui  and Kwan Lok Li }

\address{%
$^{1}$ \quad Department of Astronomy and Space Science, Chungnam National University, Daejeon 34134, Korea \\ 
$^{2}$ \quad Institute of Astronomy, National Tsing Hua University, Hsinchu 30013, Taiwan; lilirayhk@gmail.com}

\corres{Correspondence: huichungyue@gmail.com}

\abstract{The population of millisecond pulsars (MSPs) has been expanded considerably in the last decade. Not only
is their number increasing, but also various classes of them have been revealed. Among different classes of MSPs, the behaviours of
black widows and redbacks are particularly interesting. These systems consist of an MSP and
a low-mass companion star in compact binaries with an orbital period of less than a day.
In this article, we give an overview of the high energy nature of
these two classes of MSPs. Updated catalogues of black widows and redbacks are presented and their X-ray/$\gamma$-ray
properties are reviewed. Besides the overview, using the most updated eight-year {Fermi} Large Area Telescope point source
catalog, we have compared the $\gamma$-ray properties of these two MSP classes. The results suggest that the X-rays and
$\gamma$-rays observed from these MSPs originate from different mechanisms. Lastly, we will also mention the future prospects
of studying these {spider} pulsars with the novel methodologies as well as upcoming observing facilities.}


\begin{document}

\section{What Are Millisecond Pulsars?}
Rotation-powered pulsars, which act as unipolar inductors by coupling the
strong magnetic field and fast rotation, radiate at the expense of their rotational energy.~As a result, the rotation of a pulsar gradually slows down as it ages.
When the rotation becomes too slow to sustain the particles' acceleration
in their magnetospheres, the radiation beam shuts down.
In such a case, we say a pulsar is "dead". While their magnetospheric particle accelerators
have been turned off,
other emission mechanisms (e.g., accretion) can still possibly work in these dead pulsars.

\begin{figure}[H]
\centering
\includegraphics[width=15 cm]{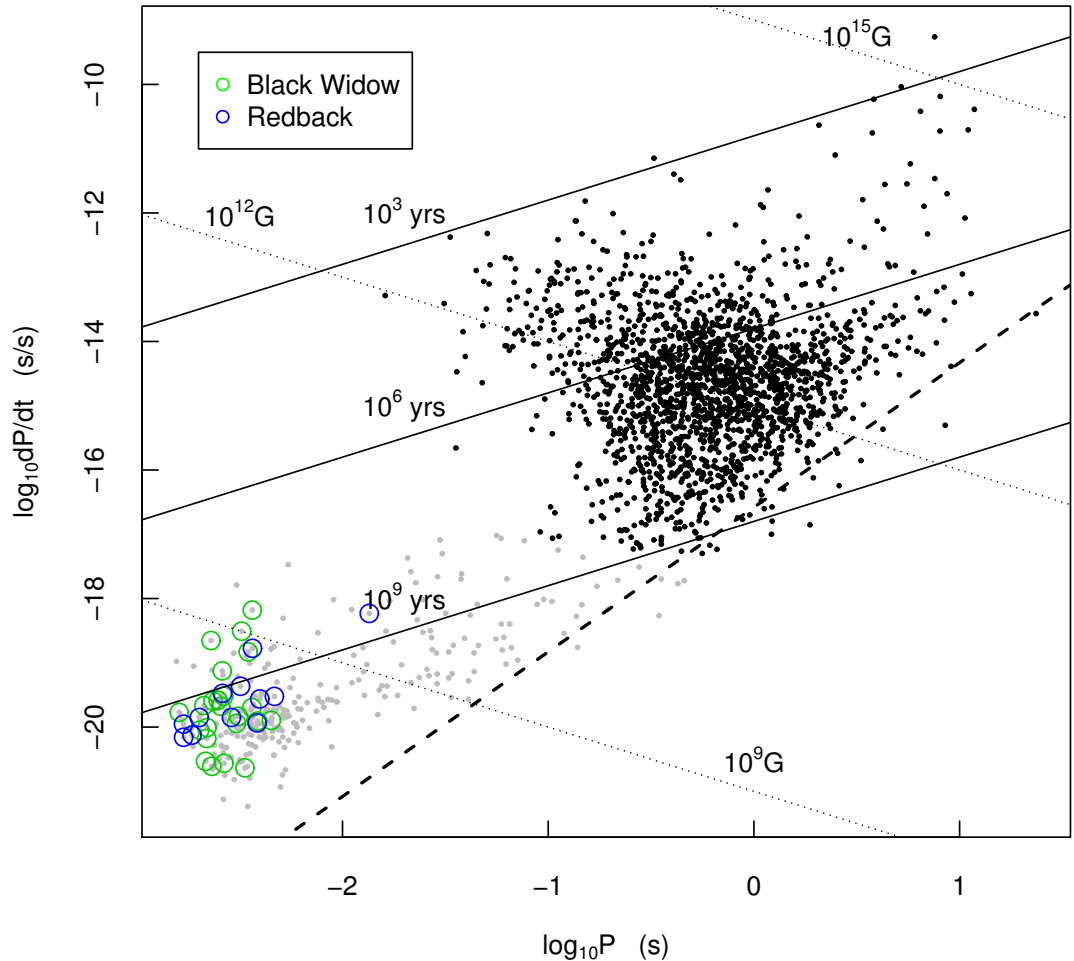}
\caption{Period--period derivative ($P$--$\dot{P}$) diagram of all currently known pulsars. The population can be divided
into two groups (black and grey dots) based on
$k$-means partitioning. The locations of black widow and redback millisecond pulsars (MSPs) in this parameter space are highlighted by the green and blue circles, respectively.
Lines of constant dipolar surface magnetic field (Equation (1)) and characteristic age (Equation (2)) are
shown. The dashed line illustrates the death line for radio pulsars by assuming a multipolar magnetic field configuration
\citep{bzhang(2000)}.\label{fig1}}
\end{figure}

By the time of writing, there are 2702 pulsars in total, including radio pulsars, radio-quiet $\gamma$-ray pulsars and
magnetars ~\citep{manchester(2005)}.~In Figure~\ref{fig1}, we show the distribution of their period $P$ and period derivative $\dot{P}$.
Applying a clustering analysis in this 2D parameter space with $k$-means partitioning,
the whole population can be divided into two parts.
The largest one is displayed as black dots in Figure~\ref{fig1}.
This partition spans a range of $P\sim$$0.02-23.5$~s and $\dot{P}\sim5\times10^{-18}-5\times10^{-10}$~s/s.
This~group includes canonical pulsars as well as magnetars.
Assuming the surface magnetic field is dipolar and the rotational energy of the pulsar goes entirely to the
dipolar radiation, one is able to estimate their surface magnetic field strength $B_{s}$ and their characteristic
age $\tau$ as:
\begin{equation}
B_{s}\simeq\sqrt{\frac{3c^{3}I}{2\pi^{2}R^{6}}P\dot{P}}\sim3\times10^{19}\sqrt{P\dot{P}}~{\rm G}
\end{equation}
\noindent and
\begin{equation}
\tau=\frac{P}{2\dot{P}}
\end{equation}
\noindent where the radius $R$ and the moment of inertia $I$ of the neutron star are taken to be
$10^{6}$~cm and \mbox{$10^{45}$~g~cm$^{2}$}~respectively.

These imply that the surface field strength and the characteristic age of this group are at the order of
$B_{s}\sim 10^{10}-10^{15}$~G and $\tau\sim 10^{3}-10^{8}$~years (see Figure~\ref{fig1}).
As a canonical pulsar ages and spins down, it moves
toward the lower right in the $P-\dot{P}$ diagram. One should note that there is an
absence of systems at the lower right region in the diagram. Such a region is a
so-called ``graveyard'' zone. The dashed line on Figure~\ref{fig1} is known as the death line and separates the neutron stars that
can sustain particle acceleration in their inner magnetospheres from those that cannot~\citep{bzhang(2000)}.
On the other hand, there is a dispersed group at the
upper right region of the $P-\dot{P}$~diagram. These objects with surface dipolar fields
$>10^{14}$~G are magnetars.


The other partition is clustered at the lower left corner of the $P-\dot{P}$ diagram,
which spans a range of $P\sim0.001-0.5$~s and $\dot{P}\sim5\times10^{-22}-10^{-17}$~s/s. The spin parameters imply these pulsars have a weak magnetic field
at the surface $B_{s}\sim10^{7}-10^{11}$~G and as old as $\tau\sim10^{9}-10^{10}$~years. These are commonly referred to as millisecond pulsars (MSPs).
They are displayed as the grey dots in Figure~\ref{fig1}.


It is a consensus that MSPs originate from evolved compact binaries.
The standard scenario for the formation of MSP is that a dead pulsar is rejuvenated
through accreting material from its binary companion~\citep{alpar1982}. During this "recycling" phase, the system
appears as a low-mass X-ray binary (LMXB). While the details of
the magnetic field decay is still a subject under discussion (e.g.,~\citep{konar2010}), the surface field of the
neutron star can be somehow buried
by the accretion. Approaching the end of this phase (on a Gyr timescale),
the neutron star will gain sufficient angular momentum and
resuscitate particle acceleration in the magnetosphere and hence the pulsed radiation again. This marks the birth of a
rotation-powered MSP.
This scenario was proposed shortly after the discovery of the first MSP PSR~B1937+21~\citep{backer1982}.
The fast rotations, weak surface magnetic fields and old ages of MSPs are all found to be consistent
with this recycling scenario~\citep{alpar1982}.

\section{{Fermi} Gamma-Ray Space Telescope---A Game Changer\label{sec2}}
\textls[-5]{Since the discovery of the first pulsar in 1967~\citep{hewish1968},
radio observations have been the major drive for the progress in pulsar astronomy for a long time.~For~hunting pulsars, a number of extensive radio surveys have been carries out. For example,
Parkes multibeam pulsar survey discovered $\sim$800~pulsars in 1997--2004 (\citep{manchester2001,lorimer2006,keith2009}). Once the sky positions of pulsars have been pin-pointed
by the radio telescopes, multiwavelength investigations of their nature can be carried out.
However, a survey of pulsars simply based on radio observations has certain limitations.~For~example,
the~ground-based radio observatories cannot provide a full sky coverage and a high cadence monitoring of their~behaviour.}


The field of pulsar survey has been changed since 2008 after the Fermi
gamma-ray space telescope was launched.
The major detector carried by Fermi is the large area telescope (LAT) which has several major improvements in
its instrumental performance in comparison with its predecessors (e.g., EGRET) (see~\citep{hui2018} for a recent review).
While the positional error attained by {EGRET} was at the order of a few degree, LAT has improved this by reducing
the error to the order of arcminute. This is particularly important for multi-wavelength follow-up investigations (see below).
Moreover, the effective area of LAT is $\sim$7000--8000~cm$^{2}$ in 1--100~GeV, which
is more than five times larger and covers a much wider energy range than its predecessors.
Furthermore, the large field of view of LAT (>2~steradian), which is more than four times larger than that of
EGRET, enables it to peform a survey of the whole sky every $\sim$3
h. This provides a much more complete sky coverage than any radio pulsar survey. Moreover, it~enables a high cadence
monitoring of the $\gamma$-ray sources almost without interruption. This turns out to be very important for studying
a special class of MSPs (see Section \ref{sec4}).

All the data acquired by
LAT are accumulated and this enables us to detect fainter sources.
5065~sources have been detected with significance $>4\sigma$ by using eight years LAT data~\citep{fermi2019}. The natures of more than one-fourth
of these $\gamma$-ray sources remain unknown. These unidentified Fermi objects provide a
large discovery space for hunting new MSPs (e.g.,~\citep{hui2015}). By imposing a set of selection criteria on
these unidentified $\gamma$-ray sources, MSP-like candidates can be selected for multiwavelength identifications. In our previous work~\citep{hui2015},
we adopted the following selection criteria: (1) High galactic latitude, (2) absence of long-term $\gamma$-ray variability and
(3) spectral shape similar to those of pulsars (i.e., a power-law with an exponential cut-off at a few GeV).
Condition 1 was imposed as MSPs are old objects and should be located far away from their
birth places (i.e., Galactic plane).
Condition 2 is imposed for discriminating them from the active galactic nuclei (AGNs).

Since the discovery of pulsar~\citep{hewish1968}, radio all-sky surveys have played a major role in advancing the pulsar astronomy
(e.g.,~\citep{manchester2001}).
However, a lot of telescope times can be wasted on the blank regions in such blind surveys.
On the other hand, the locations of the $\gamma$-ray sources uncovered by Fermi essentially provide us with a "treasure map"
for hunting pulsars.
Once the MSP-like candidates have been identified, efforts can be devoted to these sources in
searching multiwavelength counterparts within their $\gamma$-ray
positional error ellipses and identifying their nature.
As we have mentioned, the much-improved accuracy of positional determination by LAT can significantly reduce the
number of sources in the field from entering the $\gamma$-ray error ellipses by chance.
This strategy has been found to be rather efficient at discovering new MSPs. Some sources
have modulations found in X-ray and optical regimes with periodicities shorter than a day.
This is similar to the orbital modulations that have been seen in MSPs (see Sections \ref{sec3} and \ref{sec4}).
Such property makes these sources the promising MSP candidates.

\textls[-20]{{
The discovery of PSR J2339-0533 provides an example to illustrate the synergy of multiwavelength investigations of
unidentified $\gamma$-ray sources. A bright $\gamma$-ray source 0FGL J2339.8-0530 was firstly discovered in the first three months of
data collected by Fermi LAT~\citep{abdo2009}.
Within its $\gamma$-ray positional error ellipse, a bright X-ray source CXOU J233938.7-053305 was identifed as the X-ray counterpart~\citep{kong2012,2011ApJ...743L..26R}. From~the better constrained X-ray position, the optical counterpart has also been found~\citep{kong2012,2011ApJ...743L..26R}. Interestingly, modulations with a period of $\sim$4.6~h were discovered in both X-ray and optical
regimes~\citep{kong2012,2011ApJ...743L..26R}. These~properties are remarkably similar to those that have been observed in some MSP binaries which
could originate from the interaction between the pulsar wind and the ablated material from its companion and the
heating of the stellar surface of the companion by wind collision. Subsequently, a rotational period of 2.8~ms was detected in radio by the
Green Bank Telescope~\citep{ray2014}.
}}

\begin{figure}[H]
\begin{center}
\includegraphics[width=15 cm]{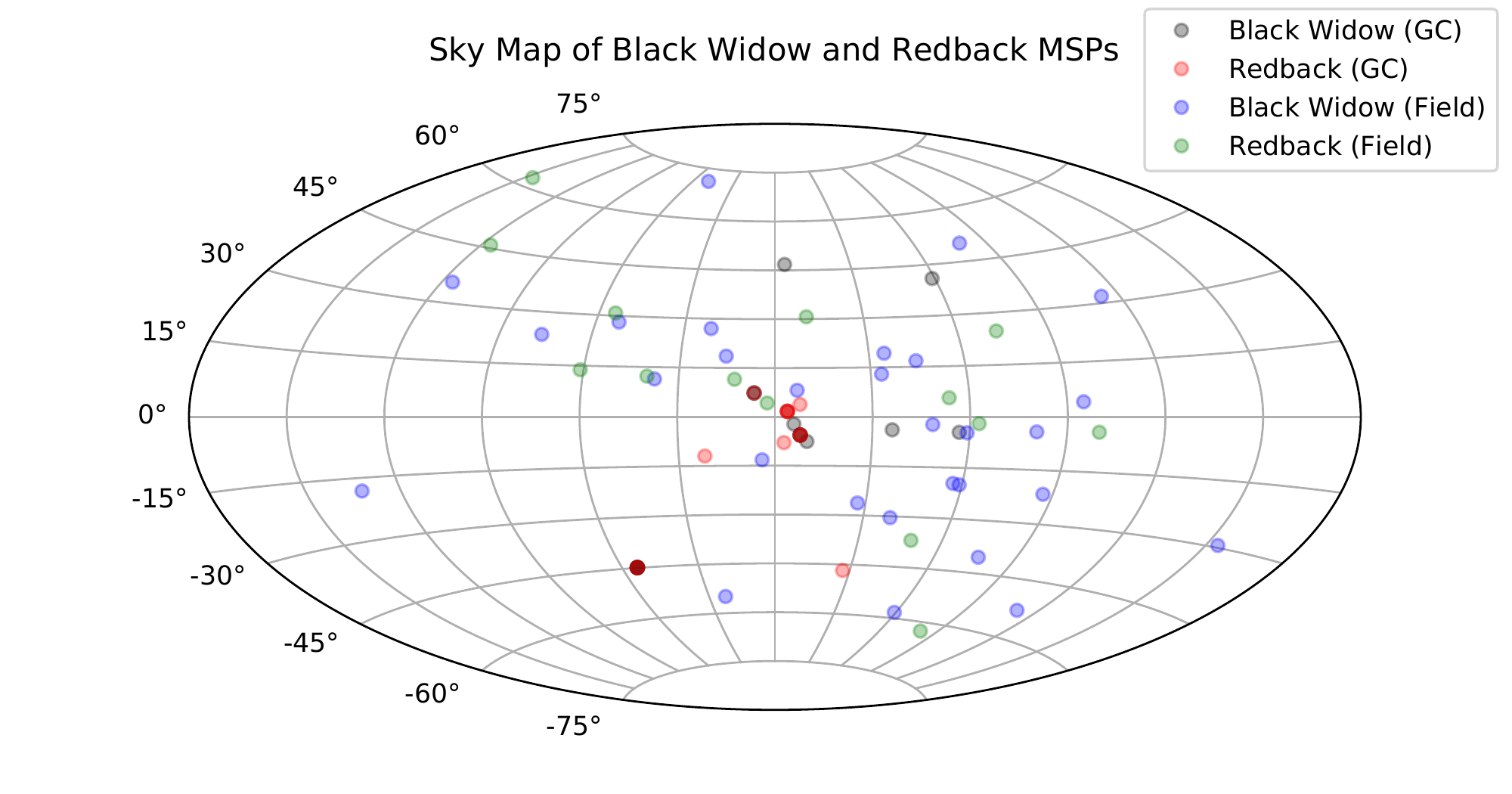}
\caption{Galactic distributions of confirmed spider pulsars. The dark red points are the results of overlapping of redbacks and black widows in
the same globular cluster.\label{fig2}}
\end{center}
\end{figure}

Among different types of MSPs, the behaviour of a group of so-called spider MSPs are the most interesting.
Spider MSPs can be divided into two classes---black widows and redbacks. {From the distribution of minimum companion mass and
the orbital period of binary MSPs in the galactic field (see Figure~1 in~\citep{robert2013}), these two classes are obviously distinct from
the other systems.}
In~\mbox{Tables~\ref{tab1} and \ref{tab2}}, we compiled the updated catalogues of all known spider MSPs in the galactic field and
globular clusters, separately.
Currently, there are 44 confirmed black widows
systems that have been detected with 27 in the galactic field and 17 reside in globular clusters. For redbacks, there are
26 in total with 14~systems confirmed in the the galactic field and 12 in globular clusters.
The distributions of black widows and redbacks in the $P-\dot{P}$ diagram are illustrated in Figure~\ref{fig1}.
In Figure~\ref{fig2}, we show their spatial distribution in our galaxy.
In the last decade, there have been many interesting high energy phenomena that have been observed from these spider pulsars.
These systems contain low-mass companion stars in tight orbits. The interactions between the MSPs and their companions
results in many interesting high energy phenomena.
In the following sections, we will give an overview of their emission properties in X-ray and $\gamma$-ray.

\begin{table}[H]
\centering
\caption{Properties of black widows (BW-F) and redbacks (RB-F) in the galactic field.\label{tab1}}
\scalebox{.85}[.85]{\begin{tabular}{lcccccccl}
\toprule
\textbf{Name} & \textbf{Type} & \textbf{Spin Period (ms)} & \textbf{Optical} & \textbf{X-ray} & \textbf{Gamma-Ray} & \textbf{Modulation}~\boldmath{$^{(1)}$} & \textbf{Pulsations}~\boldmath{$^{(2)}$} & \textbf{References}\\
\midrule
J0023+0923	&	BW-F	&	3.05	&	Y	&	Y	&	Y	&	O	& RG &	\cite{2018ApJS..235...37A,2013ApJ...769..108B,lee2018} \\
J0251+2606 & BW-F & 3.86 &   &   & Y & O  & RG & \cite{2016ApJ...819...34C,2019ApJ...883..108D} \\
J0610-2100	&	BW-F	&	3.86	&	Y	&		&	Y	&	o	& RG &	\cite{2006MNRAS.368..283B,2012ApJ...755..180P} \\
J0952-0607 & BW-F & 1.41 & Y & Y & Y & O  & RG & \cite{2017ApJ...846L..20B,2019ApJ...883..108D,2019ApJ...882..128H} \\
J1023+0038	&	RB-F	&	1.69	&	Y	&	Y	&	Y	&	OXg	& RGXO &	\cite{archibald2009,takata2014,li2014,2018RAA....18..127X,2015ApJ...807...62A,2017NatAs...1..854A} \\
J1048+2339	&	RB-F	&	4.67	&	Y	&	Y	&	Y	&	OX	& RG &	\cite{2016ApJ...819...34C,2018ApJ...866...71C,2019AA...621L...9Y} \\
J1124-3653 & BW-F & 2.41 &   & Y & Y & Ox & RG & \cite{2011AIPC.1357...40H,2019ApJ...883..108D,2014ApJ...783...69G} \\
J1227-4853	&	RB-F	&	1.69	&	Y	&	Y	&	Y	&	Og	& RGX &	\cite{2015ApJ...800L..12R,2014MNRAS.444.3004D,lee2018,2015ApJ...808...17X,2015MNRAS.449L..26P} \\
J1301+0833	&	BW-F	&	1.84	&	Y	&		&	Y	&	O	& RG &	\cite{2012arXiv1205.3089R,2014ApJ...795..115L} \\
J1302-3258	&	RB-F	&	3.77	&		&		&	Y	&		& RG &	\cite{2011AIPC.1357...40H} \\
J1311-3430	&	BW-F	&	2.56	&	Y	&	Y	&	Y	&	Og	& RG &	\cite{2012ApJ...744..105P,2012ApJ...760L..36R,lee2018,2015ApJ...804L..33X} \\
J1431-4715	&	RB-F	&	2.01	&	Y	&		&	Y	&	O	& RG &	\cite{2015MNRAS.446.4019B,2019ApJ...872...42S} \\
J1446-4701	&	BW-F	&	2.19	&		&	Y	&	Y	&		& RG &	\cite{2014MNRAS.439.1865N,lee2018} \\
J1513-2550	&	BW-F	&	2.12	&		&		&	Y	&		& RG &	\cite{san16} \\
J1544+4937	&	BW-F	&	2.16	&	Y	&		&	Y	&	O	& RG &	\cite{2013ApJ...773L..12B,2014ApJ...791L...5T} \\
J1555-2908	&	BW-F	&	1.79	&		&		&	Y	&		& RG &	\cite{wvu} \\
J1622-0315	&	RB-F	&	3.85	&	Y	&		&	Y	&	O	& RG &	\cite{san16} \\
J1628-3205	&	RB-F	&	3.21	&	Y	&	Y	&	Y	&	O	& RG &	\cite{2012arXiv1205.3089R,2014ApJ...795..115L,lee2018} \\
J1641+8049	&	BW-F	&	2.02	&	Y	&		&	Y	&		& RG &	\cite{2018ApJ...859...93L} \\
J1723-2837	&	RB-F	&	1.86	&	Y	&	Y	&		&	oX	&	R&\cite{2013ApJ...776...20C,2017ApJ...839..130K} \\
J1731-1847	&	BW-F	&	2.34	&		&	Y	&	Y	&		&	R&\cite{2014MNRAS.439.1865N,lee2018} \\
J1745+1017	&	BW-F	&	2.65	&		&		&	Y	&		& RG &	\cite{2013MNRAS.429.1633B} \\
J1805+0615	&	BW-F	&	2.13	&		&		&	Y	&		& RG &	\cite{2016ApJ...819...34C} \\
J1810+1744	&	BW-F	&	1.66	&	Y	&	Y	&	Y	&	O	& RG &	\cite{2011AIPC.1357...40H,2013ApJ...769..108B,lee2018} \\
J1816+4510	&	RB-F	&	3.19	&	Y	&	Y	&	Y	&	o	& RG &	\cite{2014ApJ...791...67S,2012ApJ...753..174K,lee2018} \\
J1832-38	&	BW-F	&	1.87	&		&		&	Y	&		&	R&\cite{wvu} \\
J1908+2105	&	RB-F	&	2.56	&	Y	&		&	Y	&		& RG &	\cite{2016ApJ...819...34C,2019ApJ...872...42S} \\
J1928+1245 & BW-F & 3.02 & Y &   &   &    & R&\cite{2019arXiv190809926P} \\
J1957+2516	&	RB-F	&	3.96	&	Y	&		&		&		&	R&\cite{2016ApJ...833..192S,2019ApJ...872...42S} \\
J1959+2048	&	BW-F	&	1.61	&	Y	&	Y	&	Y	&	OXg	& RG &	\cite{fruchter1988,2007MNRAS.379.1117R,huang2012,wu2012} \\
J2017-1614	&	BW-F	&	2.31	&	Y	&		&	Y	&	o	& RG &	\cite{san16} \\
J2047+1053	&	BW-F	&	4.29	&		&	Y	&	Y	&		& RG &	\cite{2012arXiv1205.3089R,lee2018} \\
J2051-0827	&	BW-F	&	4.51	&	Y	&	Y	&	Y	&	O	& RG &	\cite{2016MNRAS.462.1029S,1999ApJ...510L..45S,lee2018} \\
J2052+1218 & BW-F & 1.99 &   &   & Y & O  & RG & \cite{2016ApJ...819...34C,2019ApJ...883..108D} \\
J2055+3829	&	BW-F	&	2.09	&		&		&		&		&	R&\cite{2019arXiv190709778G} \\
J2115+5448	&	BW-F	&	2.60	&		&		&	Y	&		& RG &	\cite{san16} \\
J2129-0429	&	RB-F	&	7.62	&	Y	&	Y	&	Y	&	OX	& RG &	\cite{2011AIPC.1357...40H,2016ApJ...816...74B,2018MNRAS.478.3987K} \\
J2214+3000	&	BW-F	&	3.12	&	Y	&	Y	&	Y	&	O	& RG &	\cite{2018ApJS..235...37A,2014ApJ...793...78S,lee2018} \\
J2215+5135	&	RB-F	&	2.61	&	Y	&	Y	&	Y	&	oX	& RG &	\cite{2011AIPC.1357...40H,2013ApJ...769..108B,2014ApJ...783...69G} \\
J2234+0944	&	BW-F	&	3.63	&		&		&	Y	&		& RG &	\cite{2018ApJS..235...37A} \\
J2241-5236 & BW-F & 2.19 &   & Y & Y & Og & RG & \cite{2011MNRAS.414.1292K,2019ApJ...883..108D,lee2018,2018ApJ...868L...8A} \\
J2256-1024	&	BW-F	&	2.29	&	Y	&	Y	&	Y	&	OX	& RG &	\cite{2011AIPC.1357...40H,2013ApJ...769..108B,2014ApJ...783...69G} \\
J2339-0533	&	RB-F	&	2.88	&	Y	&	Y	&	Y	&	OX	& RG &	\cite{2015ApJ...807...18P,2011ApJ...743L..26R} \\
\bottomrule
\end{tabular}}

\begin{tabular}{@{}c@{}}
\multicolumn{1}{p{\textwidth -.88in}}{$^{(1)}$ O = Significant optical orbital modulation; o = possible optical orbital modulation; \mbox{X = significant X-ray orbital modulation};
x = possible X-ray orbital modulation; g = possible gamma-ray orbital modulation. $^{(2)}$ R = radio pulsations; G = gamma-ray pulsations; X = X-ray pulsations; O = optical pulsations.}
\end{tabular}

\end{table}
\unskip
\begin{table}[H]
\centering
\caption{Properties of black widows (BW-GC) and redbacks (RB-GC) in the globular clusters.\label{tab2}}
\scalebox{.83}[.83]{\begin{tabular}{lcccccccl}
\toprule
\textbf{Name} & \textbf{Type} & \textbf{Spin Period (ms)} & \textbf{Optical} & \textbf{X-ray} & \textbf{Gamma-Ray} & \textbf{Modulation}~\boldmath{$^{(1)}$} & \textbf{Pulsations}~\boldmath{$^{(2)}$} & \textbf{References}\\
\midrule
J0024-7204I	&	BW-GC	&	3.48	&		&	Y	&		&		&	R&\cite{2000ApJ...535..975C,2010arXiv1006.0335B} \\
J0024-7204J	&	BW-GC	&	2.10	&		&	Y	&		&		&	R&\cite{2000ApJ...535..975C,2010arXiv1006.0335B} \\
J0024-7204O	&	BW-GC	&	2.64	&		&	Y	&		&		&	R&\cite{2000ApJ...535..975C,2010arXiv1006.0335B} \\
J0024-7204P	&	BW-GC	&	3.64	&		&		&		&		&	R&\cite{2000ApJ...535..975C} \\
J0024-7204R	&	BW-GC	&	3.48	&		&	Y	&		&		&	R&\cite{2000ApJ...535..975C,2010arXiv1006.0335B} \\
J0024-7204V	&	RB-GC	&	4.81	&		&		&		&		&	R&\cite{2000ApJ...535..975C} \\
J0024-7204W	&	RB-GC	&	2.35	&	Y	&	Y	&		&	oX	&	R&\cite{2000ApJ...535..975C,2015ApJ...812...63C,2005ApJ...630.1029B} \\
J1518+0204C	&	BW-GC	&	2.48	&	Y	&		&		&	o	&	R&\cite{2014ApJ...795...29P} \\
J1641+3627E	&	BW-GC	&	2.49	&		&		&		&		&	R&\cite{2007ApJ...670..363H} \\
J1701-3006B	&	RB-GC	&	3.59	&	Y	&	Y	&		&	O	&	R&\cite{2012ApJ...745..109L,2008ApJ...679L.105C,2010arXiv1006.0335B} \\
J1701-3006E	&	BW-GC	&	3.23	&		&		&		&		&	R&\cite{2012ApJ...745..109L} \\
J1701-3006F	&	BW-GC	&	2.29	&		&		&		&		&	R&\cite{2012ApJ...745..109L} \\
J1740-5340	&	RB-GC	&	3.65	&	Y	&	Y	&		&	Ox	&	R&\cite{2001ApJ...561L..89D,2003AJ....125.1546K,2010arXiv1006.0335B} \\
J1748-2021D	&	RB-GC	&	13.50	&		&		&		&		&	R&\cite{2008ApJ...675..670F} \\
J1748-2446A	&	RB-GC	&	11.56	&		&		&		&		&	R&\cite{2004MNRAS.352.1439H} \\
J1748-2446O	&	BW-GC	&	1.68	&		&		&		&		&	R&\cite{2005Sci...307..892R} \\
J1748-2446P	&	RB-GC	&	1.73	&		&		&		&		&	R&\cite{2005Sci...307..892R} \\
J1748-2446ad	&	RB-GC	&	1.40	&		&		&		&		&	R&\cite{2006Sci...311.1901H} \\
J1807-2459A	&	BW-GC	&	3.06	&		&		&		&		&	R&\cite{2012ApJ...745..109L} \\
J1823-3021F	&	RB-GC	&	4.85	&		&		&		&		&	R&\cite{2012ApJ...745..109L} \\
J1824-2452G	&	BW-GC	&	5.91	&		&	Y	&		&	x	&	R&\cite{2011ApJ...730...81B,2010arXiv1006.0335B} \\
J1824-2452H	&	RB-GC	&	4.63	&	Y	&	Y	&		&	Ox	&	R&\cite{2011ApJ...730...81B,2010ApJ...725.1165P,2010arXiv1006.0335B} \\
J1824-2452I	&	RB-GC	&	3.93	&	Y	&	Y	&		&		& RX &	\cite{2011ApJ...730...81B,2013ApJ...773..122P,papitto2013} \\
J1824-2452J	&	BW-GC	&	4.04	&		&		&		&		&	R&\cite{2011ApJ...730...81B} \\
J1824-2452L	&	BW-GC	&	4.10	&		&		&		&		&	R&\cite{2011ApJ...730...81B} \\
J1836-2354A	&	BW-GC	&	3.35	&		&		&		&		&	R&\cite{2011ApJ...734...89L} \\
J1911+0102A	&	BW-GC	&	3.62	&		&		&		&		&	R&\cite{2005ApJ...621..959F} \\
J1953+1846A	&	BW-GC	&	4.89	&	Y	&	Y	&		&	O	&	R&\cite{2007ApJ...670..363H,2015ApJ...807...91C,2010arXiv1006.0335B} \\
J2140-2310A	&	RB-GC	&	11.02	&		&		&		&		&	R&\cite{2004ApJ...604..328R} \\
\bottomrule
\end{tabular}}

\begin{tabular}{@{}c@{}}
\multicolumn{1}{p{\textwidth -.88in}}{$^{(1)}$ O = Significant optical orbital modulation; o = possible optical orbital modulation; \mbox{X = significant X-ray orbital modulation};
x = possible X-ray orbital modulation; g = possible gamma-ray orbital modulation. $^{(2)}$ R = radio pulsations; G = gamma-ray pulsations; X = X-ray pulsations; O = optical pulsations}
\end{tabular}


\end{table}
\unskip

\newpage
\section{Black Widows\label{sec3}}
While the recycling scenario has been proposed for more than 30~yrs, many details of the
formation process of MSPs remain uncertain.
As mentioned before, an MSP is expected to be the end product of the evolution of a compact binary. One may be puzzled
to notice that $\sim$$30\%$ of the known MSPs in the galactic field are found to be isolated~\citep{hui2014}.
One proposed possibility to explain
their solitude is that the high energy emission from these
rejuvenated pulsars have ablated their companions~\citep{vv1988}; and the
companion stars will eventually evaporate entirely.

This proposed scenario was motivated
by the discovery of PSR~B1957+20 ( = PSR~J1959+2048)~\citep{fruchter1988}.
It is a binary that contains a 1.6~ms MSP and a very low mass companion ($M_{c}\sim0.02M_{\odot}$) in a 9.2~h orbit.
Eclipses of radio pulsations have been observed from this system during the phases called inferior conjunction
when its companion lies between the MSP and us.
Eclipses result from the absorption or scattering of the radio signals from the pulsar
by the dense ionized gas streaming off from its evaporating companion
(\citep{kluzniak1988,ruderman1989a,ruderman1989b}).
Such system is dubbed black widows because the situation is similar to a female spider that devours its
mate after mating.

While the pulsed emission is the defining characteristic of a pulsar, it only consumes a tiny fraction of the rotational
energy of the neutron star. Even with the pulsed radiation across the whole electromagnetic spectrum summed up,
it can hardly exceed a few tens of percents of a pulsar's spin-down power. Most of the rotational energy of the neutron stars is
carried away by the relativistic pulsar wind outflows. While the pulsed emission originates within the magnetosphere,
the wind region extends beyond the light cylinder. We can detect the presence of the wind through its interactions
with surroundings. For the case of PSR~B1957+20, the effect of its pulsar wind can be revealed in several~ways.

\begin{figure}[H]
\centering
\includegraphics[width=12cm]{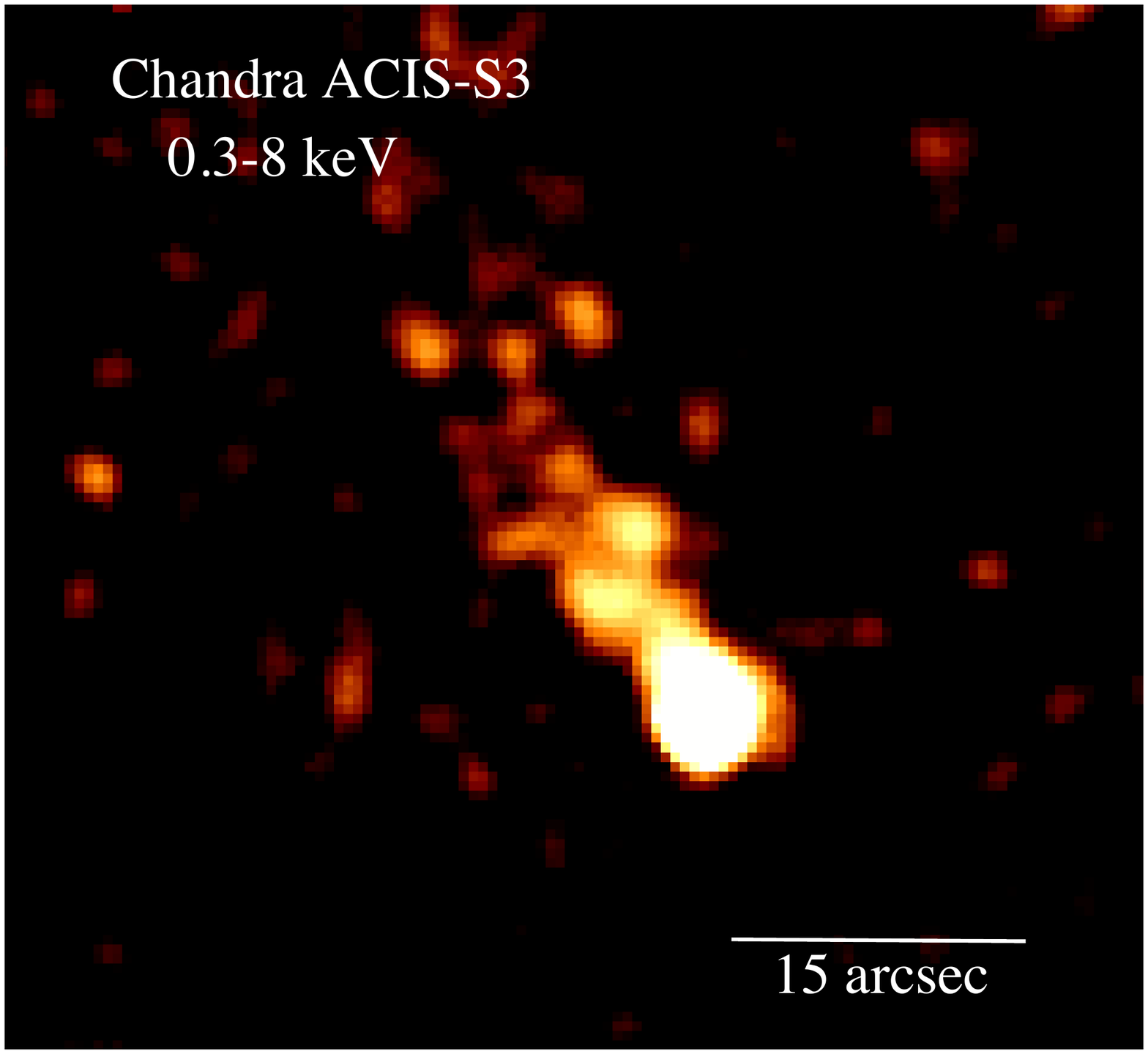}
\vspace{-2cm}
\caption{X-ray image of the bow-shock nebula associated with black widow MSP PSR~B1957+20 obtained from the public
data with an effective exposure of $\sim$165~ks as acquired by Chandra (Observation ID: 9088). This is the same data
as used in the study by \citet{huang2012}.\label{fig3}}
\end{figure}

PSR~B1957+20 has a tranverse velocity of $v\sim220$~km/s~\citep{manchester(2005)}, which is found to be supersonic
and a bow-shock nebula can be formed. The termination shock radius is determined by the balance between the wind
particles and the interstellar medium at the head of the shock.
Through the interactions with the shocked medium, the relativistic wind particles radiate synchrotron emissions as
they trail behind the pulsar's motion. This results in a cometary-tail bow-shock morphology. Figure~\ref{fig3} shows the X-ray
bow-shock nebula of PSR~B1957+20 as observed by {Chandra}, which was firstly discovered by~\citep{stappers2003}.

Systems of this kind can enable us to study the shock physics in the interstellar medium (ISM)~\citep{cheng2006}.
The observed length $l$ of the X-ray tail can be interpreted
as the distance traversed by the pulsar within the electron synchrotron cooling timescale $t_{c}$,
i.e., $l \sim v_{p}t_{c}$, where $v_{p}$ is the proper-motion velocity of the pulsar. The cooling
time in the X-ray band is $\sim$$10^{8} B_{\rm mG}^{-3/2}(h\nu_{X}/{\rm keV})^{-1/2}$~s,
where $B_{\rm mG}$ is the inferred magnetic field strength in the emitting region~\citep{cheng2006}.
Adopting the inferred tail length of \mbox{$\sim$9.7 $\times~10^{17}$~cm} at the distance of 2.5~kpc
and a proper motion velocity of 220~km~s$^{-1}$,
the synchrotron cooling timescale is estimated to be $\sim$$4.9\times 10^{4}$~yrs. This yields
a magnetic field of $B \sim 17.7 ~\upmu$G in the shock region. Considering a magnetic field strength
of $\sim 2-6 ~\mu$G in the ISM, the compression factor of the
magnetic field in the termination shock is estimated to be $\sim$3~\citep{cheng2006}.

Spatially-resolved spectroscopy can also shed light on the properties of the nebula~\citep{huang2012}.
For a synchrotron nebula powered by the MSP, a softening of the spectrum of the X-ray tail as a
function of the distance from the pulsar is expected
if synchrotron cooling of the particles injected at the termination shock is dominated. By dividing the tail
into two segments, the photon index of the segment closer to the pulsar is found to be $\Gamma\sim1.6$ and that of the
other is $\Gamma\sim2.1$~\citep{huang2012}, which is consistent with the aforementioned scenario.

Currently, there are >70 pulsar wind nebulae
detected in the X-ray~\citep{kargaltsev2013}. However, only
three of them are confirmed to be associated with MSPs (PSRs B1957+20~\citep{stappers2003}, J2124-3358~\citep{hui2006} and
J1911–1114~\citep{lee2018a}). We therefore encourage dedicated search into the archival X-ray imaging data, which can
possibly reveal unreported faint X-ray nebulae associated with MSPs.

\begin{figure}[H]
\centering
\includegraphics[width=11cm]{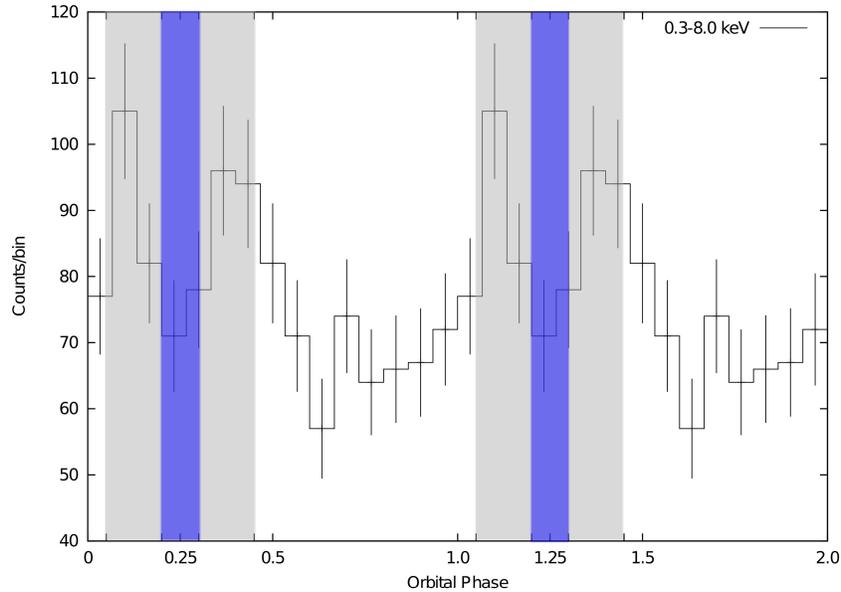}
\caption{Orbital modulation of PSR~B1957+20 in X-ray as observed by Chandra~\citep{huang2012}. The error bars
correspond to $1\sigma$ uncertainties assuming Poisson noises. The eclipses of the radio pulses occur at the orbital phases of
0.2--0.3 and 1.2--1.3 which are highlighted by the blue regions. The grey regions represent the phases for extracting the X-ray
spectrum of PSR~B1957+20 in the eclipsing region in \citet{huang2012}. Two orbital cycles are shown for clarity.\label{fig4}}
\end{figure}

The interactions between the pulsar wind and the ablated material from its companion can also form
intrabinary shock. Because of the geometry of the shock, this can lead to the modulation of high energy
emissions across the orbit. In Figure~\ref{fig4}, we show the X-ray light curve of PSR B1957+20 folded at the orbital
period~\citep{huang2012}. The observed flux is found to peak just before and just after the pulsar enters the
radio eclipse region (the blue shaded regions). This can be
interpreted as the Doppler boosting effect caused by the bulk flow in the downstream region.

The shock geometry is
controlled by the ratio of the momentum fluxes of the pulsar wind to the stellar wind.
For PSR~B1957+20, such ratio is estimated to be $\sim$10 which suggests the companion star is confined
by the pulsar wind and the shock~\citep{huang2012}. The opening angle of the cone-like shock is
$\sim$50--60$^{\circ}$, which corresponds to a separation of
$\sim$0.15 in orbital phase. Because the emission is concentrated in the forward direction
of the flow, we therefore expect that double peaks due to the Doppler boosting effect appear at the
$\sim$0.15 phase before and after the phase of the radio eclipse (i.e., inferior conjunction), which is consistent with
the observed result (see Figure~\ref{fig4})~\citep{huang2012}.

\begin{figure}[H]
\centering
\includegraphics[width=13cm]{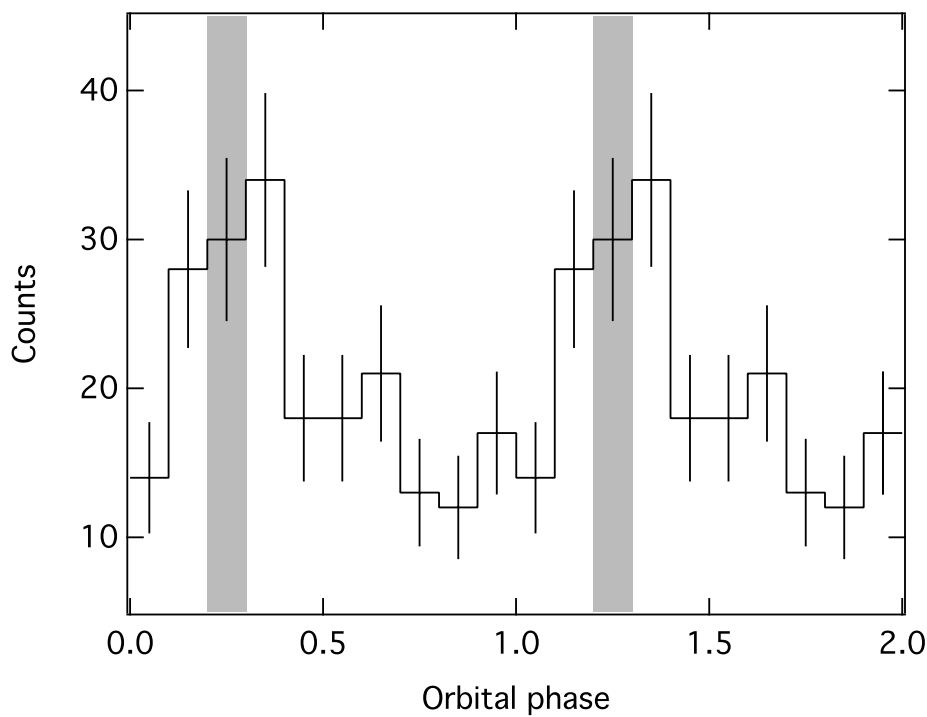}
\caption{$\gamma$-ray orbital modulation of PSR~B1957+20 observed by Fermi large area telescope (LAT). This is discovered by \citet{wu2012}.
The error bars
correspond to $1\sigma$ uncertainties assuming Poisson noises. {The shaded regions correspond to the phase of radio eclipse (i.e., 0.2--0.3
and 1.2--1.3).
Two orbital cycles are shown for clarity.}\label{fig5}}
\end{figure}

Apart from X-ray, evidence for orbital modulation of this black widow MSP is also found in a $\gamma$-ray regime.
The $\gamma$-ray emission of PSR~B1957+20 at energies larger than 2.7 GeV has been found to be modulated at the orbital period
\citep{wu2012} (see Figure~\ref{fig5}). On the other hand, the emissions below 2.7 GeV are steady and are dominated by
the pulsar emission. Figure~\ref{fig5} clearly shows that the enhanced emission above 2.7 GeV appears around
the inferior conjunction. It has been speculated that the modulated GeV emission
originates from the inverse-Compton scattering of the thermal radiation of the
companion star off the "cold" (i.e., low energy of the leptons in the co-moving frame of the plasma)
ultra-relativistic pulsar wind \citep{wu2012}.

Besides PSR~B1957+20, only two other black widow MSPs (PSR~J1311-3430~\citep{2015ApJ...804L..33X}
and PSR J2241-5236~\citep{2018ApJ...868L...8A}) have their possible
$\gamma$-ray modulations reported, which can possibly be a signature of intrabinary shock. However, their significance
is not very high, including the prototypical case of PSR~B1957+20. As more than 10 years of data have been accumulated by
Fermi LAT, one can reexamine and check these $\gamma$-ray
modulations together with the updated instrumental responses and the background model.

Currently, 27 black widows have been discovered in the galactic field (see Table~\ref{tab1}) and 17 are found in
globular clusters (Table~\ref{tab2}). The companion stars in these systems are likely to degenerate and have masses $\ll0.1M_{\odot}$
\citep{2014ApJ...783...69G,robert2013}.
For those in the galactic field, a number of them have their X-ray counterparts identified~\citep{lee2018}.
In Section~\ref{sec5}, we will provide a general discussion on the X-ray and $\gamma$-ray natures of this population as a whole.
\unskip

\section{Redbacks\label{sec4}}
Redbacks are close relatives of black widows, which are the other class of pulsar binaries that show
intense interactions between the pulsars and the companion stars.
They are named after the species of the poisonous spiders which originate from Australia.
While the range of the orbital period spanned by these systems is similar to that of black widow
MSPs ($P_{b}\leq20$~h), their companions are late-type non-degenerate stars with
masses of $M_{c}\sim0.2-0.4M_{\odot}$ which are significantly
higher than that of black widows ($M_{c}\ll0.1M_{\odot}$)~\citep{robert2013}.

Let us revisit the end stages of LMXB evolution. When a neutron star has been spun up sufficiently in the
accretion-powered phase, particle acceleration can be re-initiated in its magnetosphere and the MSP is hence turned on
\citep{alpar1982}.
However, the radio pulses cannot be observed unless the accretion disk and the dirty environment around it have been
somehow cleared. Mechanisms involve processes such as pulsar wind ablation and radiation pressure from the MSP can
possibly help to clear up the environment~\citep{alpar1982,backer1982}.
However, the exact transition between the accretion-powered LMXB and the
rotation-powered MSP has not been witnessed until a decade ago.

\begin{figure}[H]
\centering
\includegraphics[width=10cm,angle=-90]{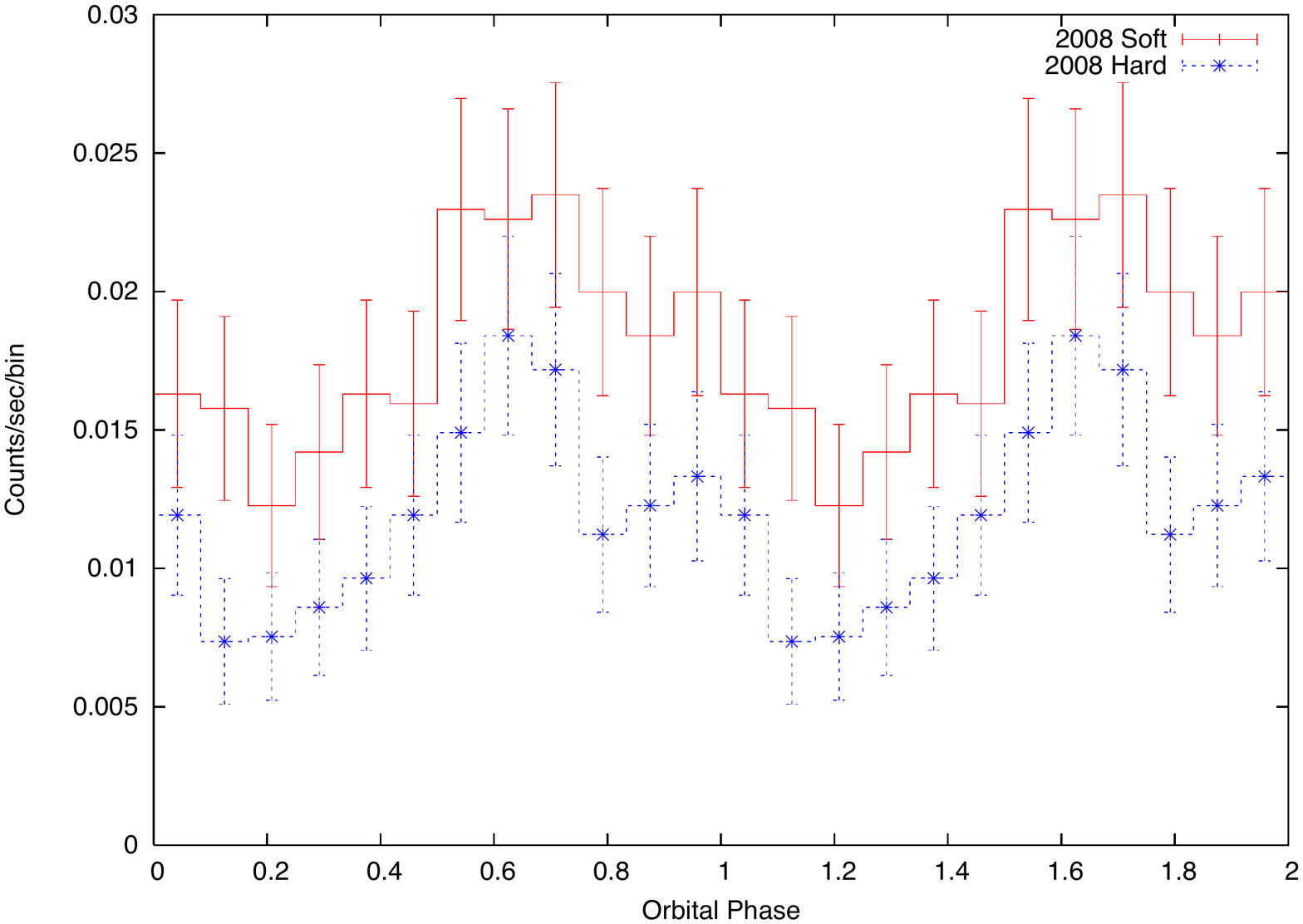}
\caption{X-ray orbital modulation of PSR J1023+0038 in the
soft (0.3--2.0 keV) and hard (2.0--10.0 keV) bands, as obtained from the
observation taken at 26 November 2008 with {XMM-Newton}~\citep{tam2010}.\label{fig6}}
\end{figure}

The first identified redback MSP, PSR~J1023+0038, was initially identified as an LMXB FIRST J102347.6+003841~\citep{homer2006}.
The source clearly showed an accretion disk before 2002~\citep{wang2009}. Since then, the disk was found to have
disappeared. Subsequently, the radio pulsations have been detected in 2007 and PSR~J1023+0038 revealed
itself as a newly born MSP~\citep{archibald2009}. It is a long-sought missing piece in the
evolutionary picture of compact X-ray binaries. Shortly after its discovery, $\gamma$-ray emission
as well as the X-ray orbital modulation (see Figure~\ref{fig6}) from PSR~J1023+0038 have been subsequently detected~\citep{tam2010}.

Interestingly, PSR~J1023+0038 shows that MSP might not be just the end point in the evolution of compact binaries.
Since 2013 late June, the radio pulsation of this system has disappeared (\citep{stappers2013,patruno2014})
and a new accretion disk has been formed as shown by the re-appearance of strong double peaked H$\alpha$
emission~\citep{halpern2013}. All this evidence indicate that this system
was switching from a rotation-powered state back to an accretion-powered state.

\begin{figure}[H]
\centering
\includegraphics[width=13cm]{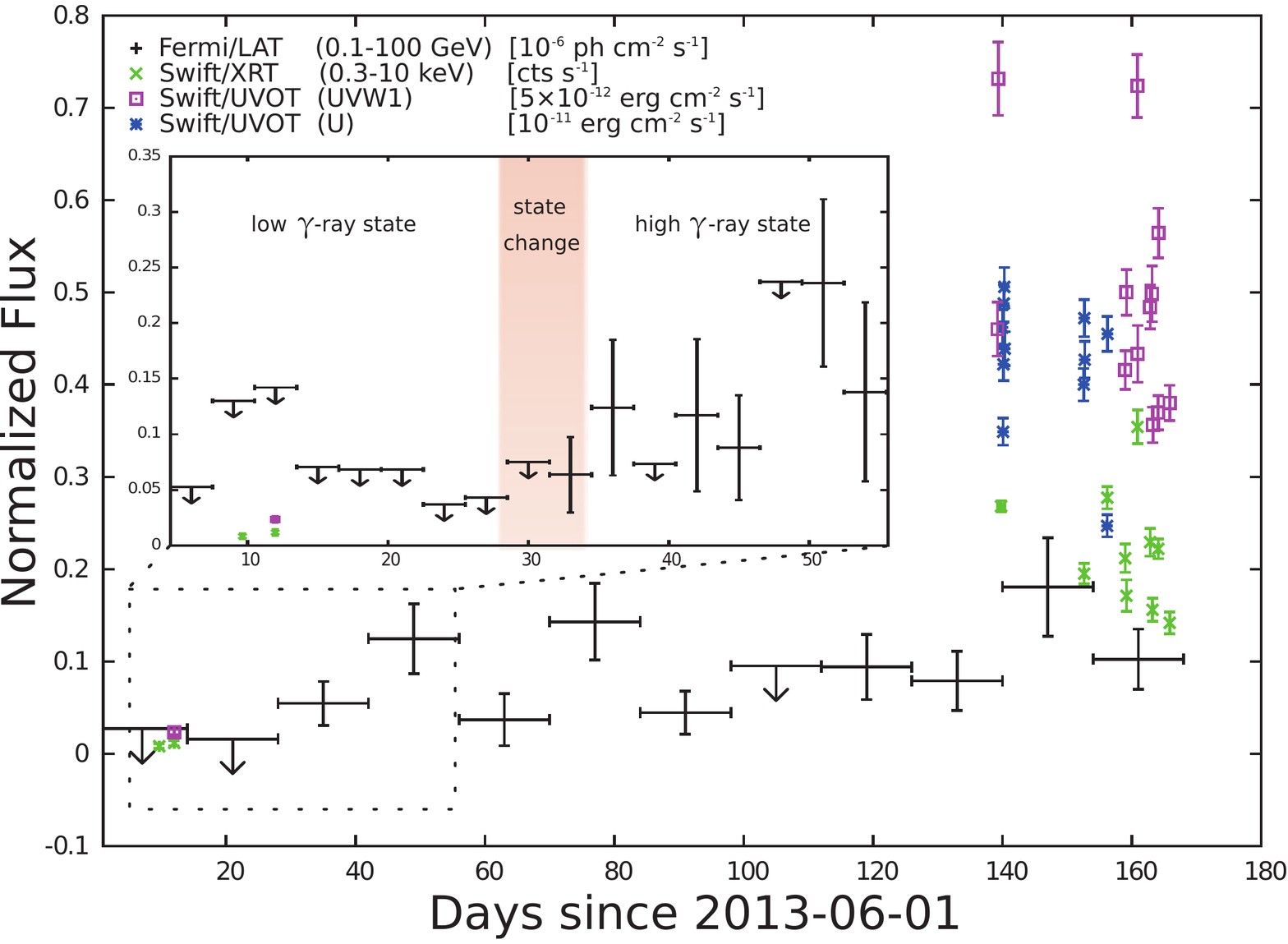}
\vspace{-4cm}
\caption{UV, X-ray and $\gamma$-ray lightcurves of PSR J1023+0038 from 1 June  2013 to 13 November 2013 are
shown together in the main panel with different flux scales for each energy band (see upper left corner for details).
On the other hand, the inset box shows the detailed evolution of the $\gamma$-ray emissions from 6 June  to 24 July.
Each data point of UV/X-ray represents an individual observation taken by {Swift}.
Each $\gamma$-ray data points in the main panel and inset corresponds to two weeks and three days, respectively.
In the cases where the detection significances is $\leq3\sigma$, upper limits at 95\% confidence are given instead~\citep{takata2014}.\label{fig7}}
\end{figure}

\textls[-15]{Changes in all other wavelengths have been found to accompany this transition (see Figure~\ref{fig7})~(\citep{takata2014,li2014}).
First, its $\gamma$-rays
suddenly brightened within a few days in 2013 June/July and have remained at a high $\gamma$-ray state.
Second, both UV and X-ray fluxes have increased by roughly an order of magnitude.
Moreover, the system does not show any X-ray orbital modulation after the transition.}

\vspace{-3cm}

\begin{figure}[H]
\centering
\includegraphics[width=7cm,angle=-90]{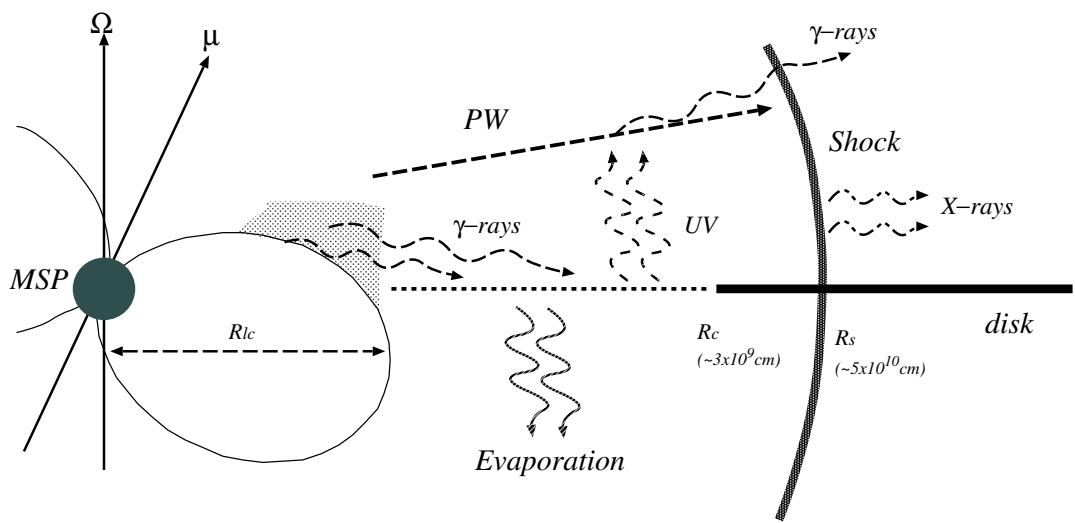}
\caption{Schematic illustration for the emission nature of
PSR~J1023+0038 after 2013 late June in different wavelengths.
The accretion disk extends beyond the light cylinder radius ($R_{lc}$).
$R_s$ is the distance to the intra-binary shock from the pulsar.
$R_{c}$ is the critical distance from the pulsar at which the $\gamma$-rays from its magnetosphere evaporate
the disk matter at $R<R_{c}\sim3\times10^{9}$~cm.  UV/Optical photons mainly originate from
the disk at $R\sim 10^{9-10}$ cm.
Shock is formed through the interaction between the pulsar wind and the stellar wind.
This produces the non-thermal X-ray emissions.
The inverse-Compton process of the cold-relativistic pulsar
wind off UV/Optical photons from the disk produces the additional $\gamma$-rays~\citep{takata2014}.\label{fig8}}
\end{figure}

Assuming the MSP is still active and the absence of radio pulsation is simply a result of absorption/scattering due to increased
local charge density, the aforementioned multiwavelength behaviour of PSR~J1023+0038 can be explained with the model
illustrated in Figure~\ref{fig8}~\citep{takata2014}. In this scenario, a new accretion disk was formed as a result of
sudden increase of the stellar wind. This can provide an explanation for all the other observed phenomena after
the transition.

The enhancement of the UV emission can be naturally accounted for by the newly formed disk.~The~increase of the $\gamma$-ray flux by an order of magnitude suggests that a new emission component emerges after
the transition.~We propose that the inverse-Compton scattering
process of the cold-relativistic pulsar wind off the optical/UV photons
from the accretion disk produces the additional $\gamma$-rays.

Before the transition in June 2013, X-ray modulation was observed from PSR J1023+0038 where
the emission originated from the intra-binary shock (\citep{tam2010,bogdanov2011}). It
has been suggested that during the MSP phase, the emission region is closer to
the companion star and its orbital variations are caused by the eclipse of the emission region by the companion star
~\citep{bogdanov2011}.
After the transition, it is possible that the size of the emission region is getting bigger
than that of before late June 2013. In~such a case, the companion star can only block a
negligible fraction of this emission and hence this explains the disappearance of the orbital X-ray variation.
It has been speculated that the increase in the mass transfer from the companion
star pushes the emission region back toward the pulsar, and more fraction
of the pulsar wind is stopped by the shock, resulting in an
increase in the X-ray emissions from the system~\citep{takata2014}.

\begin{figure}[H]
\centering
\includegraphics[width=15cm]{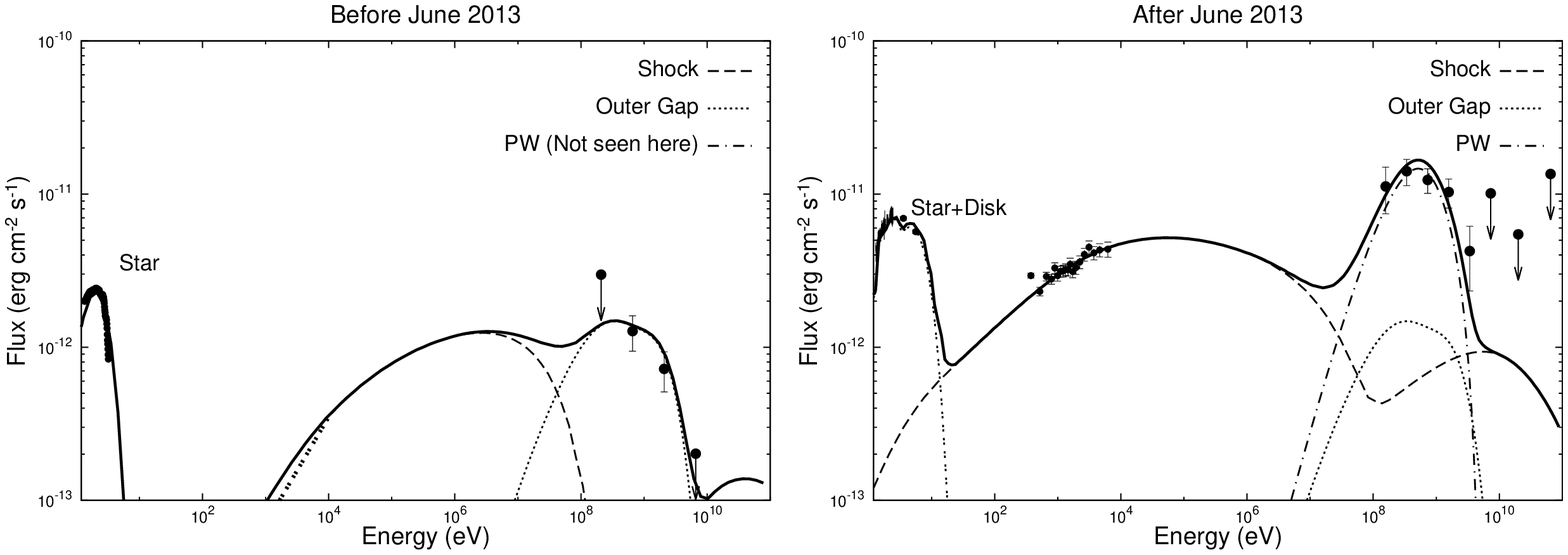}
\caption{Multi-wavelength spectral energy distributions of a PSR~J1023+0038 system before (\textbf{left}) and after (\textbf{right})
late June 2013. Calculations with a model consist of emission components from the pulsar magnetosphere (outer gap);
shock and pulsar wind (PW) are compared with the observed data before and after the transition. For further details,
please refer to~\citep{takata2014}.\label{fig9}
}
\end{figure}

In Figure~\ref{fig9}, the multiwavelength spectral energy distributions (SEDs) of PSR J1023+0038 before and after the transition in June 2013
are compared~\citep{takata2014}. Moreover, the theoretical models based on the aforementioned scenario are overlaid on the observed data.
It appears that this scenario proposed by~\citep{takata2014} can explain the results quite well.

\begin{figure}[H]
\centering
\includegraphics[width=10cm,angle=-90]{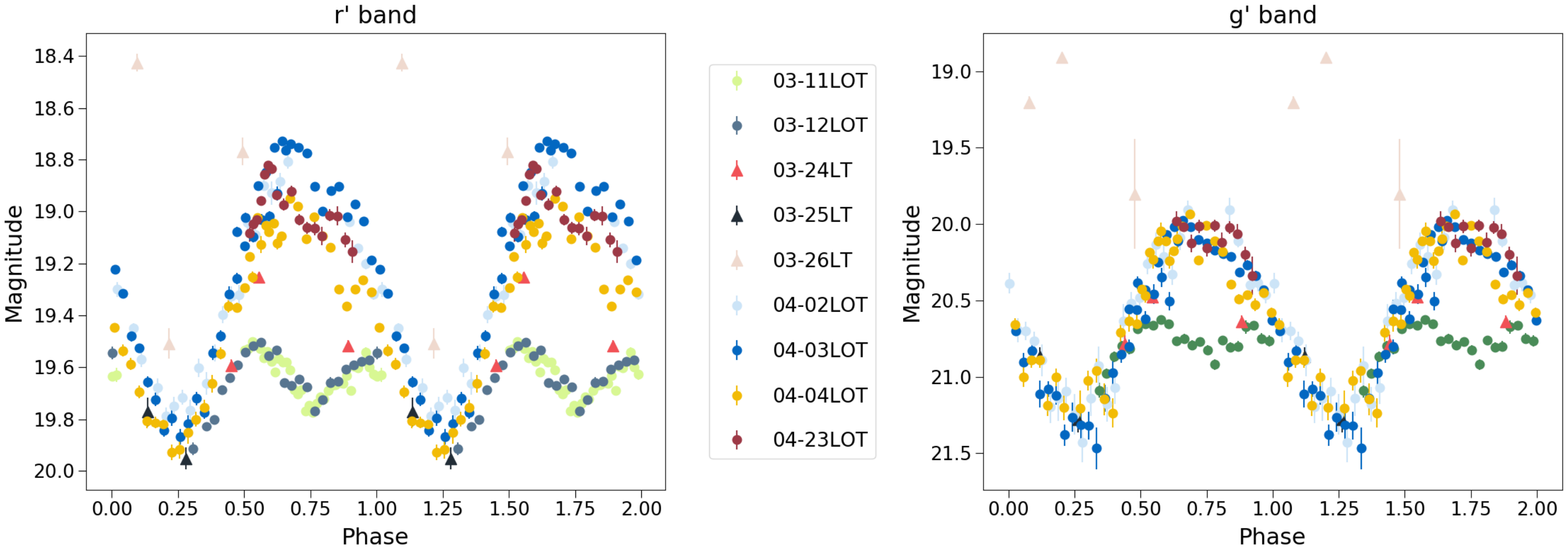}
\vspace{-2cm}
\caption{Light curves of PSR J1048+2339 companion star with $r^{'}$ and $g^{'}$ band filter,
as observed by a 1~m Lulin telescope and a 2 m Liverpool telescope
between 11 March 2018  and 23 April, folded with an orbital period of 6 h~\citep{2019AA...621L...9Y}.\label{fig10}}
\end{figure}

PSR~J1023+0038 is an archetypal examples of a growing group of transitional MSPs (tMSPs). This includes
PSR~J1824-2452I, which resides in the globular cluster M28~\citep{papitto2013}. Its X-ray counterpart is a
transient IGR J1824-2452. In 2013, it was detected as an X-ray outburst with rotational and orbital ephemeris
the same as PSR~J1824-2452I. Interestingly, after a month long X-ray burst, the system was found to be
reactivated as an active radio MSP within a few days. Another tMSP XSS J1227-4859 was previously identified
as an LMXB with positional coincidence with Fermi LAT source 1FGL J1227.9-4852/2FGL J1227.7-4853
\citep{deMartino2010}. In December 2013, a transition was accompanied with a brightness drop in opticals,
X-rays and $\gamma$-rays~\citep{bassa2013,tam2013}. Radio searches at the beginning
of 2014 eventually detected the pulses at 607~MHz from the system~\citep{roy2014}.

tMSPs are not the only redback that show dramatic changes. Very recently,
we observed another redback system PSR~J1048+2339 with the Lulin 1~m telescope in Taiwan and
2~m Liverpool telescope in La Palma~\citep{2019AA...621L...9Y}. {Its orbital modulation in the optical regime
has been found to change drastically in a timescale of less than two weeks (see Figure~\ref{fig10}).}
From the observations taken in March 2018, the optical light curve folded at the orbital period of
6~h resembles an ellipsoidal modulation of
the companion star which shows two peaks in a cycle at $\phi=0$ and $\phi=0.55$.
Ellipsoidal modulation is a consequence of
the orbital motion for a tidally distorted companion, which is commonly
seen among redback systems (e.g., PSR J2129-0429~\citep{hui2015b}). In Figure~\ref{fig10},
the minimum at $\phi=0.25$ corresponds to the inferior conjunction.

When PSR~J1048+2339 was observed again in early April 2018, its light curve was found to be completely
different from that obtained in March (see Figure~\ref{fig10}). It became a single-peaked sinusoidal profile with a maximum at
$\phi=0.65$ and a minimum at $\phi=0.25$. Furthermore, the brightness of the companion is increased by one
magnitude. All this suggests the modulation is now dominated by pulsar wind heating of the companion~\citep{2019AA...621L...9Y}.
The change from an ellipsoidal modulation to a brightened sinusoidal profile occurred in less than 14 days~\citep{2019AA...621L...9Y}.
The irradiation power of the system has been found to increase by a factor of six,
and it has been speculated that this is related to the activity of the companion star~\citep{2019AA...621L...9Y}.
The magnetic field of the companion star could play an important
role by connecting to the shock region and guides the pair
plasmas to the companion star surface.

\section{Population Analysis of Spider MSPs in High Energy\label{sec5}}
\unskip
\subsection{X-Ray Properties}
\begin{figure}[H]
\centering
\includegraphics[width=15cm]{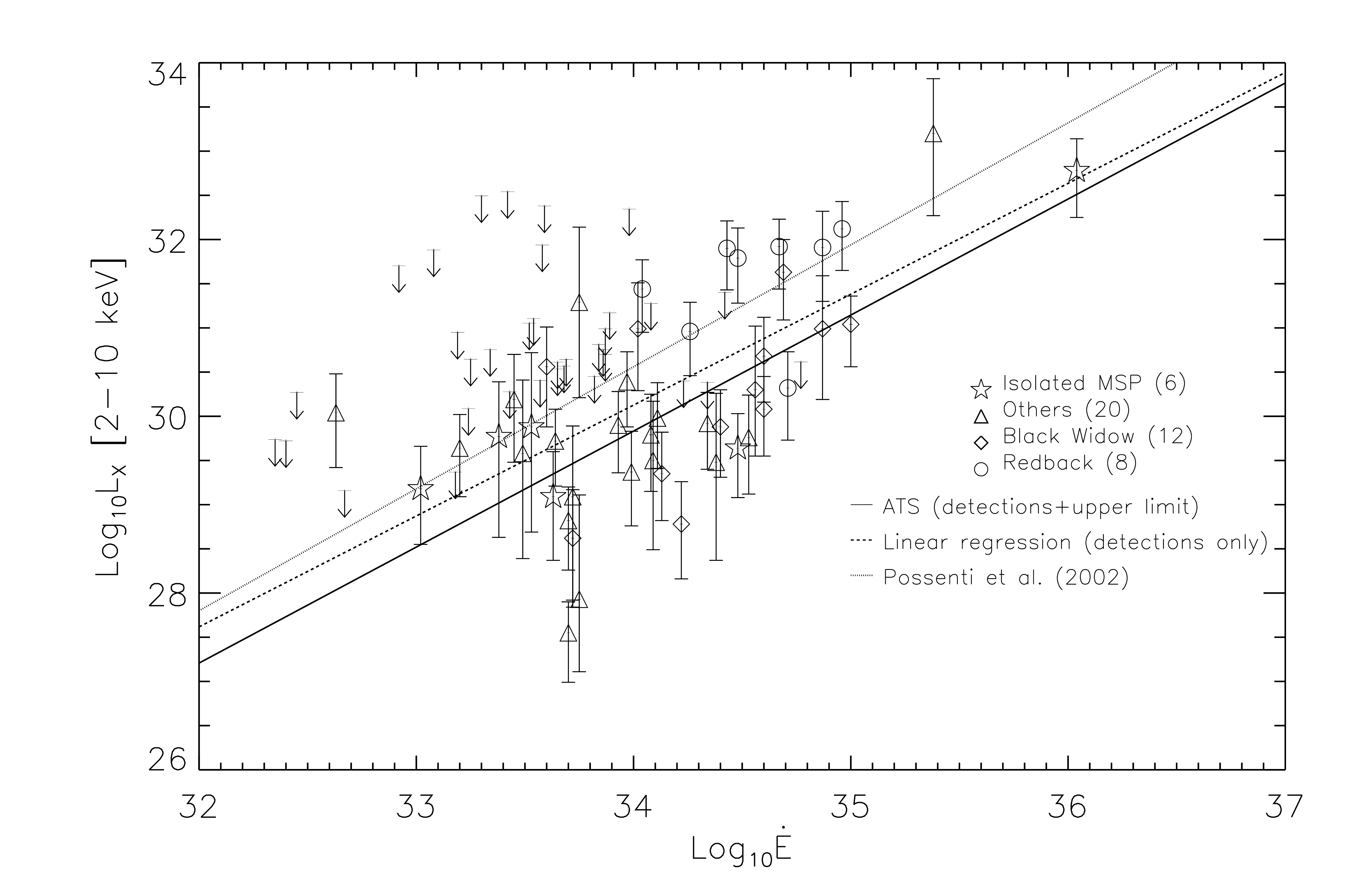}
\caption{Relation between $L_{x}$ and $\dot{E}$ for 46 MSPs of different classes which are shown as different symbols in this plot~\citep{lee2018}.
In addition, the upper-limits on $L_{x}$ for 35 MSPs are included in the sample with which \citet{lee2018} performed the survival analysis.
The solid line illustrates the Akritas--Thiel--Sen (ATS) line inferred from this censored data.
For comparison, the dashed line illustrates the result from the standard linear regression of X-ray detected MSPs. Moreover the relation reported by Possenti~et~al.~\citep{possenti2002} based on a sample of 10 MSPs is displayed as the dotted line~\citep{lee2018}.\label{fig11}}
\end{figure}

As the number of MSPs have significantly enlarged over the last decade, it is possible to
carry out a statistical analysis of their population as
a whole and examine similarities and differences among various classes.
Recently, the authors of \citet{lee2018} performed an X-ray census of all MSPs in the galactic field. By utilizing
all the available X-ray data, 47 MSPs have their X-ray counterparts detected (see Table 1 in~\citep{lee2018}).
On the other hand,  upper limits have been placed on the X-ray emission of another 36 MSPs (see Table 2 in~\citep{lee2018}).
Using this censored data, the authors of \citet{lee2018} examined the empirical relation between their X-ray luminosities $L_{x}$ and spin-down
power $\dot{E}$ (see Figure~\ref{fig11}). The best-fit relation is found to be
$L_{x}\simeq 10^{31.05}\dot{E}_{35}^{1.31}$~erg/s in 2--10~keV (i.e., the Akritas--Thiel--Sen (ATS) line in Figure~\ref{fig11}~\citep{lee2018}),
where $\dot{E}$ is the spin-down power with units of $10^{35}$~erg/s.

Concerning the spider pulsars, there are eight redbacks and 12 black widows in the X-ray selected sample adopted in the
study by \citet{lee2018}. By comparing their
physical properties with the non-spider MSPs, {while the distributions of the surface magnetic field strengths
among different classes are found to be similar}, the magnetic field strength at the light cylinder as well as the spin-down powers of
redbacks are found to be significantly higher than those of non-spider MSPs in binaries.

\begin{figure}[H]
\centering
\includegraphics[width=15cm]{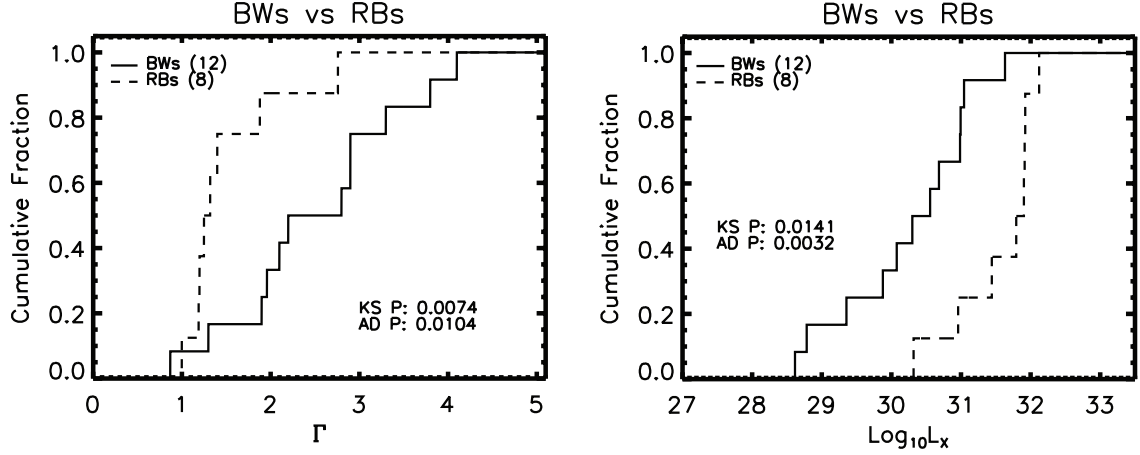}
\caption{Comparison of the effective photon indices $\Gamma$ of black-widows (BWs) and redbacks (RBs) in X-ray ({left panel}).
The comparison of the X-ray luminosities $L_{x}$ of BWs and RBs ({right panel}).
The~$p-$values resulting from the two-sample Kolmogorov--Smirnov (KS) test
and Anderson--Darling (AD) test are given in each figure, and
strongly indicate the differences between these two classes of MSPs~\citep{lee2018}.\label{fig12}}
\end{figure}

Comparing the rotational and orbital parameters of redbacks and black widows, there is no significant difference  found~\citep{lee2018}.
However, $L_{x}$ of redbacks are found to be significantly higher than those
of black widows~\citep{lee2018} (see Figure~\ref{fig12}).
Moreover, there is an indication that the X-ray emission of redbacks is harder than that of black widows~\citep{lee2018} (see Figure~\ref{fig12}).

The difference in their X-ray luminosities can be explained by the different contributions from their intrabinary
shocks~\citep{lee2018}. The shock luminosity is proportional to $L_{X}\propto \delta \dot{E}$, where $\delta$ is the fraction
  of the pulsar wind blocked by the outflow from the companion star and/or the companion star itself.
$\delta$ can be estimated by the fraction of the sky intercepted by the companion star with $\delta\sim (R_R/2a)^2$,
  where $R_R$ is the Roche-lobe radius of the companion star. Since the
  Roche-lobe radius is estimated as $R_R/a=0.462[q/(1+q)]^{1/3}$ with $q$ being  the ratio of the
  companion mass to the
  neutron star mass, the fraction can be expressed as $\delta=0.053[q/(1+q)]^{2/3}$. With the typical values of the mass ratio for
both classes of spider pulsars,
  \citet{lee2018} estimate $\delta \sim 1$\% for redbacks ($q=0.1$) and $\delta \sim 0.2$\% for  black widows ($q=0.01$).
This may explain their difference in $L_{x}$.

\subsection{$\gamma$-Ray Properties}
While the aforementioned work has investigated the X-ray properties of spider MSPs, there is no systematic
analysis of their $\gamma$-ray properties that has been reported elsewhere. For a complete review on the high
energy nature of these MSPs, we here present a statistical analysis of their $\gamma$-ray properties.

\begin{figure}[H]
\centering
\includegraphics[width=13cm]{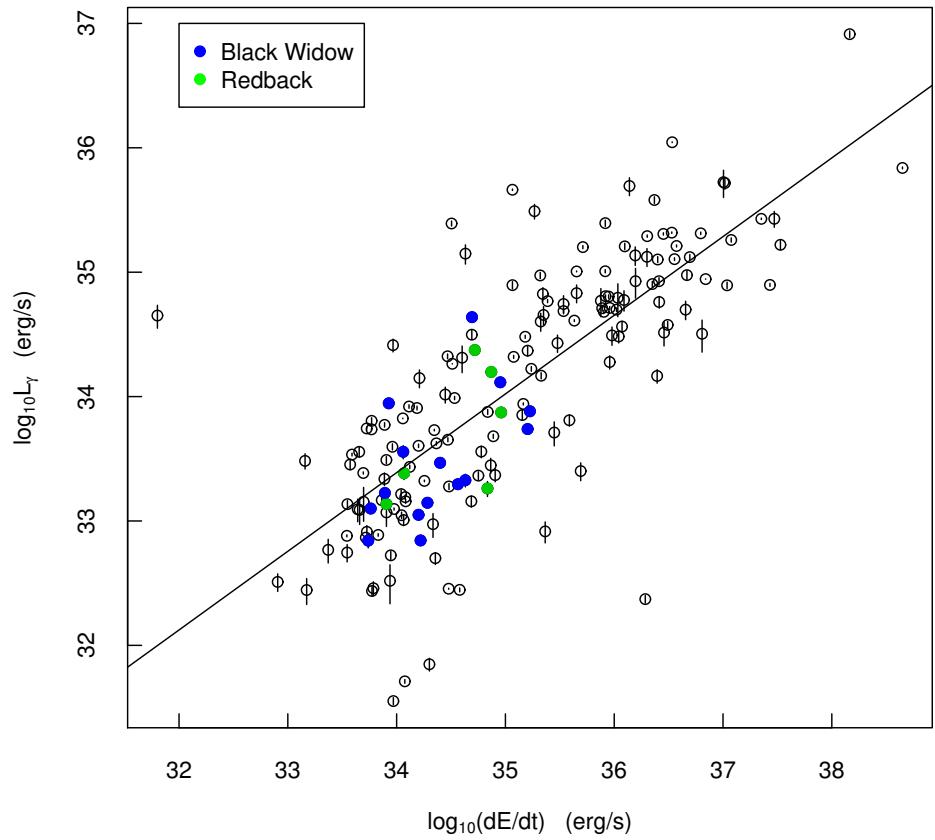}
\caption{Plot of $L_{\gamma}$ vs $\dot{E}$ for the $\gamma$-ray pulsars enlisted in the current version of the 4FGL catalog.
The~solid straight line illustrates the least square linear fit.
The locations of black widows and redbacks in this parameter space are given by blue and green symbols, respectively.\label{fig13}}
\end{figure}

First, we investigate the $L_{\gamma}-\dot{E}$ relation for the most updated sample of $\gamma$-ray pulsars,
where $L_{\gamma}$ is the $\gamma$-ray luminosity in 0.1--100~GeV.
We make use of the Fermi LAT 8-Year Point Source Catalog (4FGL) of the version updated on 29 May 2019
\citep{fermi2019}.
Among 5065 sources detected at a significance above $4\sigma$, 239 of them are classified as pulsars.
For computing their $L_{\gamma}$ and $\dot{E}$, we need the estimates of their distance as well as $\dot{P}$.
Searching for such properties of the 4FGL pulsars from the ATNF catalog~\citep{manchester(2005)}, we obtain a
sample of 169 pulsars with their $L_{\gamma}$ and $\dot{E}$ plotted in Figure~\ref{fig13}. By fitting a linear model
to $\log L_{\gamma}$ and $\log\dot{E}$, we obtain a best-fit relation of

\begin{equation}
\log L_{\gamma} = (0.63\pm0.05)\log\dot{E}+(11.87\pm 1.78)
\end{equation}
\noindent with the uncertainties estimated by bootstraping. We plot this relation as a straight line in Figure~\ref{fig13}.
The data points for the spider pulsars are highlighted by the coloured symbols. We found they follow
the general $L_{\gamma}-\dot{E}$ trend inferred from the entire $\gamma$-ray population.

{Since the luminosities and the spectral properties of black widows and redbacks in X-ray regimes
are found to be rather different~\citep{lee2018}}, it is instructive to compare the corresponding properties of these two classes in $\gamma$-ray.
The $\gamma$-ray spectrum of a pulsar is typically characterized by an exponentially cutoff power law with the
following form:

\begin{equation}
\frac{dN}{dE}=K\left(\frac{E}{E_{0}}\right)^{-\Gamma}\exp\left(a\left(E_{0}^{b}-E^{b}\right)\right)
\end{equation}

\noindent where $E_{0}$ is a chosen energy scale expressed in MeV, $K$ is the prefactor which describes the flux density of the pulsar,
$\Gamma$ is the photon index which describes the low energy spectral slope,
$a$ and $b$ are the exponential factor and and the exponential index which describe
the cutoff at high energy. In the adopted 4FGL catalog, except for the six brightest $\gamma$-ray pulsars which were
fitted with Equation (4) with $b$ as free parameter, all the other pulsars are fitted with this function with $b$ fixed at 2/3
\citep{fermi2019}.

\begin{figure}[H]
\centering
\includegraphics[width=14.5cm]{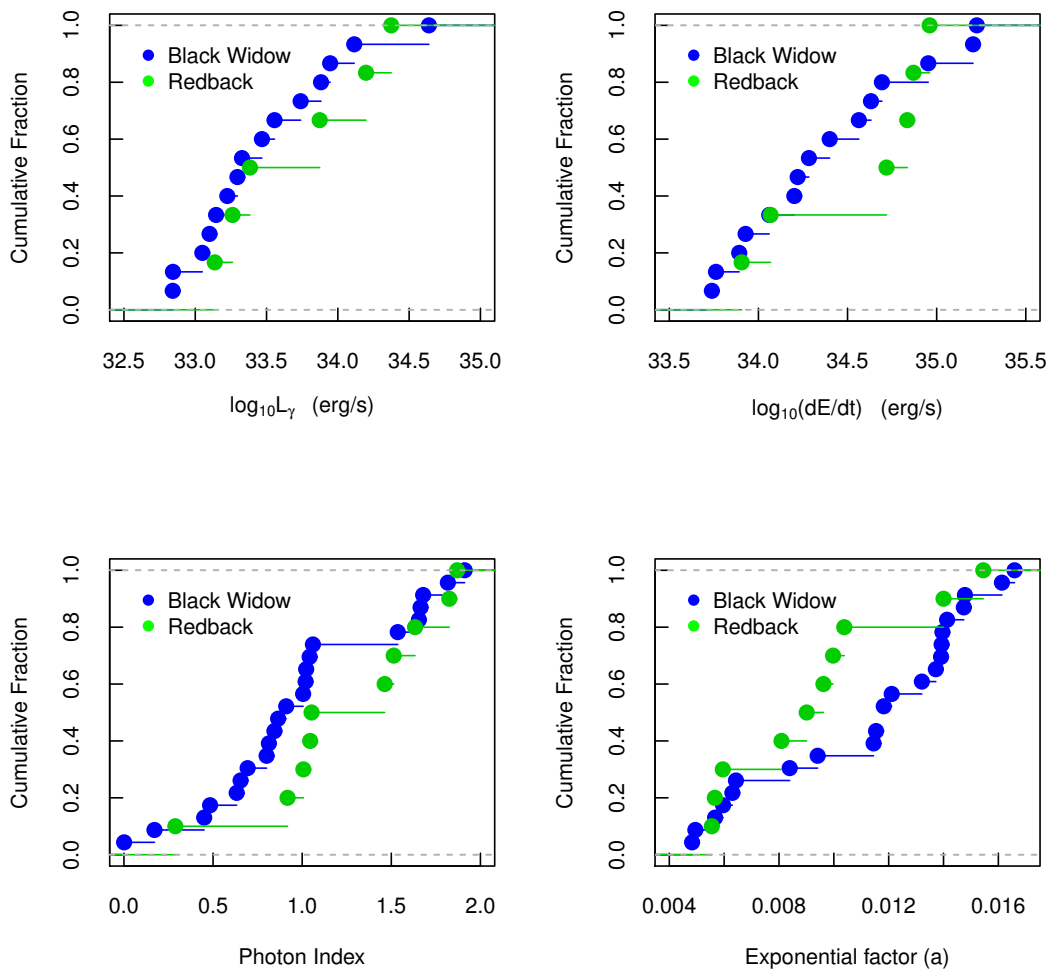}
\caption{Comparisons of the step-wise empirical cumulative distributions of
$L_{\gamma}$ (upper-left panel), $\dot{E}$~(upper-right panel), $\gamma$-ray photon index
(lower-left panel) and $\gamma$-ray exponential factor $a$ (lower-right panel) of black-widows and redbacks in $\gamma$-rays. \label{fig14}}
\end{figure}

In Figure~\ref{fig14}, we construct the cumulative distribution functions for the $\gamma$-ray properties of black widows (blue symbols) and
the redbacks (green symbols). We first compare $L_{\gamma}$ of these two classes of spider MSPs (top left panel of Figure~\ref{fig14}).
In contrast to the case for X-rays, the distributions of $L_{\gamma}$ of these two groups appear to be similar to each other.
Applying a two-samples Anderson--Darling (A-D) test to their distribution, which is the same method adopted by
\citep{lee2018} in comparing the X-ray properties of different classes of MSPs, we obtain a $p$-value of 0.6 and hence we conclude that there is no difference
in the $\gamma$-ray luminosity distributions between black widows and redbacks.
We have also compared the $\dot{E}$ of this $\gamma$-ray selected sample (top right panel of Figure~\ref{fig14}) and we also do not find
any difference between black widows and redbacks.

In the lower-left and lower-right panels of Figure~\ref{fig14}, we compare the distributions of $\Gamma$ and $a$ between the groups of
23 black widows and 10 redbacks from the 4FGL catalog. While these plots show some differences in their $\gamma$-ray spectral
properties, A-D tests yield the $p$-values of 0.16 and 0.30 in comparing their distributions of $\Gamma$ and $a$, respectively.
Hence, we conclude that the $\gamma$-ray spectral properties of black widows and redbacks are consistent with each other.

In this analysis, we do not find any significant difference in the $\gamma$-ray properties between black widows and redbacks.
The different contributions from the intrabinary shocks in these two MSP classes, which lead to their differences in the X-ray,
do not result in any notable difference in the $\gamma$-ray regime. This suggests that the X-rays and $\gamma$-rays from these spider
MSPs originated from different mechanisms.
we conclude that the $\gamma$-ray emission of the spider pulsars are dominated by their magnetospheric~radiation.
\unskip

\section{Future Prospects}
We have seen that a vast of dramatic high energy phenomena have been observed from the spider MSPs. Here we would like to
give a wish list for what kind of investigations we can do and what we can expect to learn from these targets in the near future.

\begin{figure}[H]
\centering
\includegraphics[width=14cm]{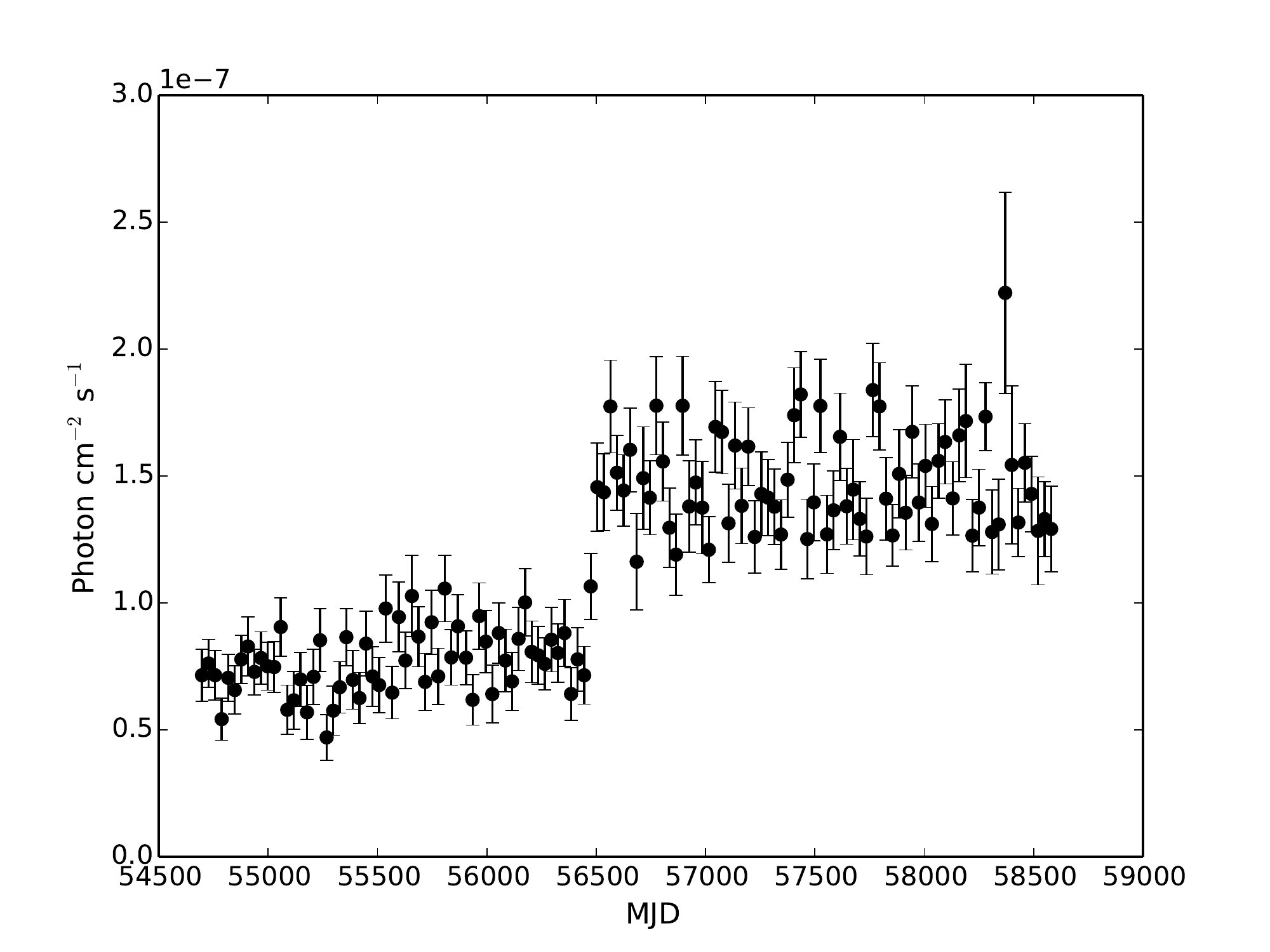}
\caption{A long-term $\gamma$-ray light curve of PSR~J1023+0038
as observed by Fermi LAT at energies $>100$~MeV from MJD 54697 (August 2008) to MJD 58580 (April 2019).\label{fig15}}
\end{figure}

{1. Catch the changes in the act:}
PSR~J1023+0038 is a classic example of a redback and demonstrates the swings between
accretion-powered state and rotation-powered
state. The emission properties of the system have been found to change drastically along with the transition (Figures \ref{fig7}--\ref{fig9}).
Studying~these variabilities can allow better understanding of the evolution of an LMXB in the recycling phase. However,
it is impossible to catch such a transition without a long-term monitoring campaign. Almost all the spider pulsars are
$\gamma$-ray emitters. Therefore, the continuous all-sky surveying mode of Fermi LAT provides us with the ideal
instrument to monitor their behaviours. In Figure~\ref{fig15}, we show the long-term $\gamma$-ray light curve of PSR~J1023+0038
as observed by Fermi LAT. The transition in June 2013 (MJD~56450) can be clearly noted. Once a similar jump
in the $\gamma$-ray flux has been spotted from any redbacks,
a fast response X-ray/UV follow-up with the {Neil Gehrels Swift} Observatory can hence be~triggered.

{2. Globular clusters vs. Galactic field:}
In Section 5, we highlighted the results from the population analysis of the X-ray selected MSPs in the galactic
field. We mentioned that the X-ray emission from the redbacks in the galactic field are apparently
brighter and harder than the other classes of MSPs, including black widows (Figure~\ref{fig12}). In Tables \ref{tab1} and \ref{tab2}, we have shown that the
current sample sizes of the spider pulsars in the galactic field and in globular clusters are comparable.
The formation processes of MSPs in globular clusters is dynamical~\citep{hui2010},
which can be very different from those in galactic field.
It is instructive to compare the rotational, orbital and emission properties between these
two populations. As the properties of redbacks in the galactic field are found to be different from the other
classes, it is interesting to see if their counterparts in globular clusters also share similar behaviours.
One can also compare the redbacks/black widows in these two populations to explore if the dynamical formation processes
in globular clusters can result in any difference.

{3. Spider hunting:}
Thanks to Fermi, the MSP population has been significantly enlarged through the synergy of picking MSP-like
candidates from the unidentified $\gamma$-ray sources and the multiwavelength identification campaign. In our previous MSP-like candidates selection, we adopt a simple
set of selection criteria which was based on our current knowledge of MSPs (see Section \ref{sec2}). However, the accuracy of picking the
right candidates by this conventional method is unlikely to be optimal because the patterns in the data that define an
MSP can be overlooked by human investigators. To improve the efficiency and the accuracy of the MSP identification
campaign, machine learning techniques should be employed in $\gamma$-ray source classification.

We note that some studies have used such techniques to classify $\gamma$-ray sources (e.g.,~\citep{parkinson2016}).
However, since the features for building the classification model were selected manually in these works,
it is unlikely that the power of machine learning has been fully exploited. Recently, we have developed
a scheme by coupling the classification algorithms with automatic feature extraction methods~\citep{leung2017}.
With this scheme, one can improve the prediction accuracy and pick the MSP-like $\gamma$-ray sources with higher confidence. Using the
known $\gamma$-ray selected MSPs for testing our algorithm, we can attain an accuracy $>95\%$ in classifying MSPs while
all the previous attempts can only achieve at most $\sim$90\%.
Therefore, we expect that the multiwavelength follow-up observations of the targets selected by this method can result in
more spider pulsars.

Very recently, a new X-ray mission {eROSITA} was launched, on 13 July 2019.\footnote{\url{https://www.mpe.mpg.de/eROSITA}.}
{eROSITA} is equipped with seven identical
Wolter-I mirror modules. Each module consists of 54 nested mirror shells. It is going to perform an all-sky survey in 0.3--10 keV
with unprecedented sensitivity. Therefore, it will provide a full X-ray coverage of all unidentified $\gamma$-ray sources. On the other hand,
the upcoming large synoptic survey telescope (LSST)\footnote{\url{https://www.lsst.org}.}
will have an 8.4~m mirror with an exceptionally wide field of $3.5^{\circ}$.
It is capable of surveying the entire observable sky in the optical regime from its site in only three nights. It can provide
deep multi-epoch data for the periodicity searches which enable us to identify the companions of the MSP-like objects.

In Table~\ref{tab3}, we summarize the properties of a number of promising spider pulsar candidates. They~were identified by the
periodicities detected in X-ray/optical which resemble those orbital modulations found in redbacks/black widows.~Three of these targets, 3FGL~J0427.9$-$6704, FL8Y~J1109.8$-$6500 and 3FGL~J1544.6$-$1125, have shown stepwise jumps
in their long-term X-ray light curves that are similar to the transitions between accretion-powered and rotation-powered states
that have seen in PSR~J1023+0038. Furthermore, H$\alpha$ emissions have been detected from these three sources, which indicates the
possibility of accretion. This makes them the candidates of tMSPs. Follow-up observations of them are strongly encouraged.
\footnote{We would like to point out that FL8Y~J1109.8$-$6500 was enlisted in the preliminary catalogue with eight years of LAT data. However, it is
excluded in the most recent version~\citep{fermi2019}.}.

\begin{table}[H]
\centering
\caption{List of promising spider pulsar candidates.\label{tab3}}
\scalebox{.9}[.9]{\begin{tabular}{ccccccl}
\toprule
\textbf{Name} & \textbf{Orbital Period} & \textbf{Mass Function }& \textbf{X-ray Detection} & \textbf{Accretion} & \textbf{Modulation} & \textbf{References} \\
 & \textbf{(day)} & \textbf{(}\boldmath{$M_{\odot}$}\textbf{)} & \textbf{(Y/N)} & \textbf{(Y/N)} & &  \\
\midrule
3FGL~J0212.1+5320 & 0.870 & 0.88 & Y & N & Ox & \cite{2016ApJ...833..143L,2017MNRAS.465.4602L} \\
3FGL~J0427.9$-$6704 & 0.367 & 0.96 & Y & Y & OXg & \cite{2016ApJ...831...89S} \\
1FGL~J0523.5$-$2529 & 0.688 & 0.49 & Y & N & Og & \cite{2014ApJ...788L..27S,2014ApJ...795...88X} \\
3FGL~J0744.1$-$2523 & 0.115 &  & N & N & O & \cite{2017MNRAS.470..466S} \\
3FGL~J0802.3$-$5610 & 0.416 &  & Y & N & O & \cite{2017MNRAS.470..466S} \\
3FGL~J0838.8$-$2829 & 0.215 & 0.69 & Y & N & O & \cite{2017ApJ...844..150H} \\
2FGL~J0846.0+2820 & 8.133 & 0.14 & N & Y? & o & \cite{2017ApJ...851...31S} \\
3FGL~J0954.8$-$3948 & 0.387 & 0.81 & Y & N & Ox & \cite{2018ApJ...863..194L} \\
FL8Y~J1109.8$-$6500 &  &  & Y & Y & & \cite{2019AA...622A.211C} \\
3FGL~J1544.6$-$1125 & 0.242 & 0.0015 & Y & Y & O & \cite{2015ApJ...803L..27B,2017ApJ...849...21B} \\
2FGL~J1653.6$-$0159 & 0.052 & 1.60 & Y & N & Ox & \cite{2014ApJ...793L..20R,2014ApJ...794L..22K} \\
3FGL~J2039.6$-$5618 & 0.228 & 0.80 & Y & N & OXg & \cite{2015ApJ...812L..24R,2015ApJ...814...88S,2019ApJ...872...42S,2018ApJ...867...90N} \\
\bottomrule
\end{tabular}}
\begin{tabular}{@{}c@{}}
\multicolumn{1}{p{\textwidth -.88in}}{\footnotesize O = Significant optical orbital modulation; o = Possible optical orbital modulation; X = Significant X-ray orbital modulation;
x = Possible X-ray orbital modulation; g = Possible gamma-ray orbital modulation. }
\end{tabular}

\end{table}

{4. Push to very high energy regime:} In a very high energy regime (VHE; >100~GeV), neither the pulsed emission nor
the pulsar wind nebula has been detected from any MSP so far.
However, the optical and X-ray modulations observed from spider MSPs
indicate the presence of ablation and heating of their companions.
The interaction of the pulsar and stellar winds can create a shocked region in which particles can be accelerated
\citep{arons1993}. Such accelerated electrons can possibly Comptonize stellar radiation from the companion
to the TeV energies. This process can produce orbital modulated VHE $\gamma$-rays.

On the other hand, black widow PSR B1957+20 has a bow-shock X-ray nebula (See Figure~\ref{fig3}).
Assuming the observed non-thermal X-ray emission up to $\sim10$ keV is produced in a
synchrotron process, this implies the magnetic field of the nebula of
the order of $\mu$G and the presence of high energy electrons with energies up to the order of a hundred TeV~\citep{cheng2006}.
These electrons can also possibly Comptonize the ambient soft photon fields and result in VHE $\gamma$-rays
\citep{bednarek2013}. It is interesting to note that the limiting fluxes in the TeV regime placed by MAGIC are very
close to the theoretically predicted values (see Figure~8 in~\citep{ahnen2017}).

Cherenkov telescope array (CTA) is the next generation VHE observatory which
consists of more than 100 telescopes located in both the northern and southern hemispheres.
Upon completion of the construction of the full array,  CTA will have a much larger collecting area,
wider energy coverage and a larger field-of-view than any existing VHE-observing facility.
With its unprecedented performance at such high energies, it is not unreasonable to speculate that a number of spider
pulsars can be detected in the VHE regime. This will open a new widow for us to study these MSPs and
help us to further constrain the acceleration processes of leptons within the binaries and their surrounding.

\vspace{6pt}



\authorcontributions{The two authors contributed equally.} 

\funding{This research was funded by the National Research Foundation of Korea through grant numbers 2016R1A5A1013277 and
2019R1F1A1062071.}

\acknowledgments{C.Y.H. is supported by the National Research Foundation of Korea through grants 2016R1A5A1013277 and
2019R1F1A1062071. K.L.L. is supported by the Ministry of Science and Technology of Taiwan through grant 108-2112-M-007-025-MY3.}

\conflictsofinterest{The authors declare no conflict of interest.} 






\reftitle{References}



\begin{thebibliography}{999}

\bibitem[]{manchester(2005)}
Manchester, R.N.; Hobbs, G.B.; Teoh, A.; Hobbs, M. The Australia Telescope National Facility Pulsar Catalogue. {\em Astron. J.} {\bf 2005}, {\em 129}, 1993--2006.
\bibitem[]{bzhang(2000)} Zhang, B.; Harding, A.K.; Muslimov, A.G. Radio Pulsar Death Line Revisited: Is PSR J2144-3933 Anomalous? {\em \apj} {\bf 2000}, {\em 531}, L135--L138.
\bibitem[]{alpar1982}
Alpar, M.A.; Cheng, A.F.; Ruderman, M.A.; Shaham, J. A new class of radio pulsars. {\em  Nature} {\bf 1982}, {\em 300}, 728--730.
\bibitem[]{konar2010}
Konar, S. The magnetic fields of millisecond pulsars in globular clusters. {\em Mon. Not. R.  Astron. Soc.} {\bf 2010}, {\em 409}, 259--268.
\bibitem[]{backer1982}
Backer, D.C.; Kulkarni, S.R.; Heiles, C.; Davis, M.M.; Goss, W.M. A millisecond pulsar. {\em Nature} {\bf 1982}, {\em 300}, 615--618.
\bibitem[]{hewish1968}
Hewish, A.; Bell, S.J.; Pilkington, J.D.H.; Scott, P.F.; Collins, R.A. Observation of a Rapidly Pulsating Radio Source. {\em Nature} {\bf 1968}, {\em 217}, 709--713.
\bibitem[]{lorimer2006}
Lorimer, D.R.; Faulkner, A.J.; Lyne, A.G.; Manchester, R.N.; Kramer, M.; McLaughlin, M.A.; Hobbs, G.; Possenti, A.; Stairs, I.H.; Camilo, F.; et al. The Parkes Multibeam Pulsar Survey---VI. Discovery and timing of 142 pulsars and a Galactic population analysis. {\em Mon. Not. R.  Astron. Soc.} {\bf 2006}, {\em 372}, 777--800.
\bibitem[]{keith2009}
Keith, M.J.; Eatough, R.P.; Lyne, A.G.; Kramer, M.; Possenti, A.; Camilo, F.; Manchester, R.N. Discovery of 28 pulsars using new techniques for sorting pulsar candidates. {\em Mon. Not. R.  Astron. Soc.} {\bf 2009} 395, 837--846.
\bibitem[]{manchester2001}
Manchester, R.N.; Lyne, A.G.; Camilo, F.; Bell, J.F.; Kaspi, V.M.; D'Amico, N.; McKay, N.P.F.; Crawford, F.; Stairs, I.H.; Possenti, A.; et al. The Parkes multi-beam pulsar survey---I. Observing and data analysis systems, discovery and timing of 100 pulsars. {\em Mon. Not. R.  Astron. Soc.} {\bf 2001},  {\em 328}, 17--35. 
\bibitem{hui2018}
Hui, C.Y. A Golden Decade of Gamma-Ray Pulsar Astronomy. {\emph{J. Korean Astron. Soc.}} {\bf 2018}, {\em 51}, 171--183
\bibitem[]{fermi2019}
Abdollahi, S.; Acero, F.; Ackermannn, M.; Ajello, M.; Atwood, W.B.; Axelsson, M.; Baldini, L.; Ballet, J.; Barbiellini, G.; Bastieri, D.; et al. Fermi Large Area Telescope Fourth Source Catalog. \emph{arXiv} {\bf 2019}, arXiv:1902.10045v3.
\bibitem[]{hui2015}
Hui, C.Y.; Park, S.M.; Hu, C.P.; Lin, L.C.C.; Li, K.L.; Kong, A.K.H.; Tam, P.H.T.; Takata, J.; Cheng, K.S.; Jin, R.; et al. Searches for Millisecond Pulsar Candidates among the Unidentified Fermi Objects. {\em  \apj} {\bf 2015}, {\em 809}, 68.
\bibitem[]{abdo2009}
Abdo, A.A.; Ackermann, M.; Ajello, M.; Atwood, W.B.; Axelsson, M.; Baldini, L.; Ballet, J.; Band, D.L.; Barbiellini, G.; Bastieri, D.; et al. Fermi/Large Area Telescope Bright Gamma-Ray Source List. {\emph{Astrophys. J. Suppl. Ser.}} {\bf 2009}, {\em 183}, 46--66.
\bibitem[{Romani} and {Shaw}(2011)]{2011ApJ...743L..26R}
{Romani}, R.W.; {Shaw}, M.S.
\newblock {The Orbit and Companion of Probable {\ensuremath{\gamma}}-Ray Pulsar
  J2339-0533}.
\newblock {\em Astrophys. J. Lett.} {\bf 2011}, {\em 743},~L26,
\newblock
  doi:{\changeurlcolor{black}\href{https://doi.org/10.1088/2041-8205/743/2/L26}{\detokenize{10.1088/2041-8205/743/2/L26}}}
\bibitem[]{kong2012}
Kong, A.K.H.; Huang, R.H.H.; Cheng, K.S.; Takata, J.; Yatsu, Y.; Cheung, C.C.; Donato, D.; Lin, L.C.C.; Kataoka, J.; Takahashi, Y.; et al. Discovery of an Unidentified Fermi Object as a Black Widow-like Millisecond Pulsar. {\em \apjl} {\bf 2012}, {\em 747}, L3.
\bibitem[]{ray2014}
Ray, P.S.; Belfiore, A.M.; Saz Parkinson, P.; Polisensky, E.; Ransom, S.M.; Romani, R.W.; Hessels, J.; Razzano, M.; Bhattacharyya, B.; Roy, J.; et al. Discovery of the radio and gamma-ray pulsar PSR J2339-0533 associated with the Fermi LAT bright source 0FGL J2339.8-0530. In \emph{American~Astronomical Society,  AAS Meeting \#223}; id.140.07; American~Astronomical Society: Washington, DC, USA, 2014.
\bibitem[]{robert2013}
Roberts M.S.E. Surrounded by spiders! New black widows and redbacks in the galactic field. {\em Proc. Int. Astron. Union} {\bf 2013}, {\em 291}, 127--132.
\bibitem[{Arzoumanian} \em{et~al.}(2018){Arzoumanian}, {Brazier},
  {Burke-Spolaor}, {Chamberlin}, {Chatterjee}, {Christy}, {Cordes}, {Cornish},
  {Crawford}, {Thankful Cromartie}, {Crowter}, {DeCesar}, {Demorest}, {Dolch},
  {Ellis}, {Ferdman}, {Ferrara}, {Fonseca}, {Garver-Daniels}, {Gentile},
  {Halmrast}, {Huerta}, {Jenet}, {Jessup}, {Jones}, {Jones}, {Kaplan}, {Lam},
  {Lazio}, {Levin}, {Lommen}, {Lorimer}, {Luo}, {Lynch}, {Madison}, {Matthews},
  {McLaughlin}, {McWilliams}, {Mingarelli}, {Ng}, {Nice}, {Pennucci}, {Ransom},
  {Ray}, {Siemens}, {Simon}, {Spiewak}, {Stairs}, {Stinebring}, {Stovall},
  {Swiggum}, {Taylor}, {Vallisneri}, {van Haasteren}, {Vigeland}, {Zhu}, and
  {NANOGrav Collaboration}]{2018ApJS..235...37A}
{Arzoumanian}, Z.; {Brazier}, A.; {Burke-Spolaor}, S.; {Chamberlin}, S.;
  {Chatterjee}, S.; {Christy}, B.; {Cordes}, J.M.; {Cornish}, N.J.; {Crawford},
  F.; {Thankful Cromartie}, H.; et al.
\newblock {The NANOGrav 11-year Data Set: High-precision Timing of 45
  Millisecond Pulsars}.
\newblock {\em  Astrophys. J.  Suppl. Ser.} {\bf 2018}, {\em 235},~37, doi:10.3847/1538-4365/aab5b0.

\bibitem[{Breton} \em{et~al.}(2013){Breton}, {van Kerkwijk}, {Roberts},
  {Hessels}, {Camilo}, {McLaughlin}, {Ransom}, {Ray}, and
  {Stairs}]{2013ApJ...769..108B}
{Breton}, R.P.; {van Kerkwijk}, M.H.; {Roberts}, M.S.E.; {Hessels}, J.W.T.;
  {Camilo}, F.; {McLaughlin}, M.A.; {Ransom}, S.M.; {Ray}, P.S.; {Stairs}, I.H.
\newblock {Discovery of the Optical Counterparts to Four Energetic Fermi
  Millisecond Pulsars}.
\newblock {\em  Astrophys. J.  Suppl. Ser.} {\bf 2013}, {\em 769},~108,
\newblock
  doi:{\changeurlcolor{black}\href{https://doi.org/10.1088/0004-637X/769/2/108}{\detokenize{10.1088/0004-637X/769/2/108}}}.

\bibitem[Lee et al. (2018)]{lee2018} Lee, J.; Hui, C.Y.; Takata, J.; Kong, A.K.H.; Tam, P.H.T.; Cheng, K.S. X-ray Census of Millisecond Pulsars in the Galactic Field. {\em \apj} {\bf 2018}, {\em 864}, 23

\bibitem[{Cromartie} \em{et~al.}(2016){Cromartie}, {Camilo}, {Kerr}, {Deneva},
  {Ransom}, {Ray}, {Ferrara}, {Michelson}, and {Wood}]{2016ApJ...819...34C}
{Cromartie}, H.T.; {Camilo}, F.; {Kerr}, M.; {Deneva}, J.S.; {Ransom}, S.M.;
  {Ray}, P.S.; {Ferrara}, E.C.; {Michelson}, P.F.; {Wood}, K.S.
\newblock {Six New Millisecond Pulsars from Arecibo Searches of Fermi Gamma-Ray
  Sources}.
\newblock {\em  Astrophys. J.} {\bf 2016}, {\em 819},~34,
\newblock
  doi:{\changeurlcolor{black}\href{https://doi.org/10.3847/0004-637X/819/1/34}{\detokenize{10.3847/0004-637X/819/1/34}}}.

\bibitem[Draghis et al.(2019)]{2019ApJ...883..108D} Draghis, P.; Romani, R.W.; Filippenko, A.V.; Brink, T.G.; Zheng, W.; Halpern, J.P.; Camilo, F. Multiband Optical Light Curves of Black-widow Pulsars. {\em  Astrophys. J.} {\bf 2019}, {\em 883}, 108.

\bibitem[{Burgay} \em{et~al.}(2006){Burgay}, {Joshi}, {D'Amico}, {Possenti},
  {Lyne}, {Manchester}, {McLaughlin}, {Kramer}, {Camilo}, and
  {Freire}]{2006MNRAS.368..283B}
{Burgay}, M.; {Joshi}, B.C.; {D'Amico}, N.; {Possenti}, A.; {Lyne}, A.G.;
  {Manchester}, R.N.; {McLaughlin}, M.A.; {Kramer}, M.; {Camilo}, F.; {Freire},
  P.C.C.
\newblock {The Parkes High-Latitude pulsar survey}.
\newblock {\em Mon. Not. R. Astron. Soc.} {\bf 2006}, {\em 368},~283--292,
\newblock
  doi:{\changeurlcolor{black}\href{https://doi.org/10.1111/j.1365-2966.2006.10100.x}{\detokenize{10.1111/j.1365-2966.2006.10100.x}}}.

\bibitem[{Pallanca} \em{et~al.}(2012){Pallanca}, {Mignani}, {Dalessandro},
  {Ferraro}, {Lanzoni}, {Possenti}, {Burgay}, and {Sabbi}]{2012ApJ...755..180P}
{Pallanca}, C.; {Mignani}, R.P.; {Dalessandro}, E.; {Ferraro}, F.R.; {Lanzoni},
  B.; {Possenti}, A.; {Burgay}, M.; {Sabbi}, E.
\newblock {The~Identification of the Optical Companion to the Binary
  Millisecond Pulsar J0610-2100 in the Galactic Field}.
\newblock {\em  Astrophys. J.} {\bf 2012}, {\em 755},~180,
\newblock
  doi:{\changeurlcolor{black}\href{https://doi.org/10.1088/0004-637X/755/2/180}{\detokenize{10.1088/0004-637X/755/2/180}}}.

\bibitem[Bassa et al.(2017)]{2017ApJ...846L..20B} Bassa, C.G.; Pleunis, Z.; Hessels, J.W.T.; Ferrara, E.C.; Breton, R.P.; Gusinskaia, N.V.; Kondratiev, V.I.; Sanidas, S.; Nieder, L.; Clark, C.J.; et al. LOFAR Discovery of the Fastest-spinning Millisecond Pulsar in the Galactic Field. {\em  Astrophys. J. Lett.} {\bf 2017}, {\em 846}, L20.

\bibitem[Ho et al.(2019)]{2019ApJ...882..128H} Ho, W.C.G.; Heinke, C.O.; Chugunov, A.I. XMM-Newton Detection and Spectrum of the Second Fastest Spinning Pulsar PSR J0952{\ensuremath{-}}0607. {\em  Astrophys. J.} {\bf 2019}, {\em 882}, 128.


\bibitem[]{archibald2009} Archibald, A.M.; Stairs, I.H.; Ransom, S.M.; Kaspi, V.M.; Kondratiev, V.I.; Lorimer, D.R.; McLaughlin, M.A.; Boyles, J.; Hessels, J.W.T.; Lynch, R.; et al. A Radio Pulsar/X-ray Binary Link. {\em Science} {\bf 2009}, {\em 324}, 1411--1414.

\bibitem[]{takata2014} Takata, J.; Li, K.L.; Leung, G.C.K.; Kong, A.K.H.; Tam, P.H.T.; Hui, C.Y.; Wu, E.M.H.; Xing, Y.; Cao, Y.; Tang, S.; et al. Multi-wavelength Emissions from the Millisecond Pulsar Binary PSR J1023+0038 during an Accretion Active State. {\em \apj} {\bf 2014}, {\em 785}, 131.

\bibitem[]{li2014} Li, K.L.; Kong, A.K.H.; Takata, J.; Cheng, K.S.; Tam, P.H.T.; Hui, C.Y.; Jin, R. NuSTAR Observations and Broadband Spectral Energy Distribution Modeling of the Millisecond Pulsar Binary PSR J1023+0038. {\em \apj} {\bf 2014}, {\em 797}, 111.

\bibitem[{Xing} \em{et~al.}(2018){Xing}, {Wang}, and
  {Takata}]{2018RAA....18..127X}
{Xing}, Y.; {Wang}, Z.X.; {Takata}, J.
\newblock {Possible modulated {\ensuremath{\gamma}}-ray emission from the
  transitional millisecond pulsar binary PSR J1023+0038}.
\newblock {\em Res. Astron. Astrophys.} {\bf 2018}, {\em
  18},~127,
\newblock
  doi:{\changeurlcolor{black}\href{https://doi.org/10.1088/1674-4527/18/10/127}{\detokenize{10.1088/1674-4527/18/10/127}}}.

\bibitem[]{2015ApJ...807...62A} Archibald, A.M.; Bogdanov, S.; Patruno, A.; Hessels, J.W.T.; Deller, A.T.; Bassa, C.; Janssen, G.H.; Kaspi, V.M.; Lyne, A.G.; Stappers, B.W.; et al. Accretion-powered Pulsations in an Apparently Quiescent Neutron Star Binary. {\em \apj} {\bf 2015}, {\em 807}, 62.

\bibitem[]{2017NatAs...1..854A} Ambrosino, F.; Papitto, A.; Stella, L.; Meddi, F.; Cretaro, P.; Burderi, L.; Di Salvo, T.; Israel, G.L.; Ghedina, A.; Di Fabrizio, L.; et al. Optical pulsations from a transitional millisecond pulsar. {\em Nat. Astron.} {\bf 2017}, {\em 1}, 854.

\bibitem[{Cho} \em{et~al.}(2018){Cho}, {Halpern}, and
  {Bogdanov}]{2018ApJ...866...71C}
{Cho}, P.B.; {Halpern}, J.P.; {Bogdanov}, S.
\newblock {Variable Heating and Flaring of Three Redback Millisecond Pulsar
  Companions}.
\newblock {\em  Astrophys. J.} {\bf 2018}, {\em 866},~71,
\newblock
  doi:{\changeurlcolor{black}\href{https://doi.org/10.3847/1538-4357/aade92}{\detokenize{10.3847/1538-4357/aade92}}}.

\bibitem[{Yap} \em{et~al.}(2019){Yap}, {Li}, {Kong}, {Takata}, {Lee}, and
  {Hui}]{2019AA...621L...9Y}
{Yap}, Y.X.; {Li}, K.L.; {Kong}, A.K.H.; {Takata}, J.; {Lee}, J.; {Hui}, C.Y.
\newblock {Face changing companion of the redback millisecond pulsar PSR
  J1048+2339}.
\newblock {\em Astron. Astrophys.} {\bf 2019}, {\em 621},~L9,
\newblock
  doi:{\changeurlcolor{black}\href{https://doi.org/10.1051/0004-6361/201834545}{\detokenize{10.1051/0004-6361/201834545}}}.

\bibitem[{Hessels} \em{et~al.}(2011){Hessels}, {Roberts}, {McLaughlin}, {Ray},
  {Bangale}, {Ransom}, {Kerr}, {Camilo}, and {Decesar}]{2011AIPC.1357...40H}
{Hessels}, J.W.T.; {Roberts}, M.S.E.; {McLaughlin}, M.A.; {Ray}, P.S.;
  {Bangale}, P.; {Ransom}, S.M.; {Kerr}, M.; {Camilo},~F.; {Decesar}, M.E.
\newblock {A 350-MHz GBT Survey of 50 Faint Fermi {\ensuremath{\gamma}}-ray
  Sources for Radio Millisecond Pulsars}.
\newblock  In {\em American Institute of Physics Conference Series}; {Burgay}, M.,
  {D'Amico}, N., {Esposito}, P., {Pellizzoni}, A., {Possenti}, A., Eds.; American Institute of Physics: New York, NY, USA, 2011; Volume 1357,  pp. 40--43,
\newblock
  doi:{\changeurlcolor{black}\href{https://doi.org/10.1063/1.3615072}{\detokenize{10.1063/1.3615072}}}.

\bibitem[{Gentile} \em{et~al.}(2014){Gentile}, {Roberts}, {McLaughlin},
  {Camilo}, {Hessels}, {Kerr}, {Ransom}, {Ray}, and
  {Stairs}]{2014ApJ...783...69G}
{Gentile}, P.A.; {Roberts}, M.S.E.; {McLaughlin}, M.A.; {Camilo}, F.;
  {Hessels}, J.W.T.; {Kerr}, M.; {Ransom}, S.M.; {Ray}, P.S.; {Stairs}, I.H.
\newblock {X-ray Observations of Black Widow Pulsars}.
\newblock {\em  Astrophys. J.} {\bf 2014}, {\em 783},~69,
\newblock
  doi:{\changeurlcolor{black}\href{https://doi.org/10.1088/0004-637X/783/2/69}{\detokenize{10.1088/0004-637X/783/2/69}}}.

\bibitem[{Roy} \em{et~al.}(2015){Roy}, {Ray}, {Bhattacharyya}, {Stappers},
  {Chengalur}, {Deneva}, {Camilo}, {Johnson}, {Wolff}, {Hessels}, {Bassa},
  {Keane}, {Ferrara}, {Harding}, and {Wood}]{2015ApJ...800L..12R}
{Roy}, J.; {Ray}, P.S.; {Bhattacharyya}, B.; {Stappers}, B.; {Chengalur}, J.N.;
  {Deneva}, J.; {Camilo}, F.; {Johnson}, T.J.; {Wolff},~M.; {Hessels}, J.W.T.;
 et al.
\newblock {Discovery of PSR J1227-4853: A Transition from a Low-mass X-ray
  Binary to a Redback Millisecond Pulsar}.
\newblock {\em Astrophys. J. Lett.} {\bf 2015}, {\em 800},~L12,
\newblock
  doi:{\changeurlcolor{black}\href{https://doi.org/10.1088/2041-8205/800/1/L12}{\detokenize{10.1088/2041-8205/800/1/L12}}}.

\bibitem[{de Martino} \em{et~al.}(2014){de Martino}, {Casares}, {Mason},
  {Buckley}, {Kotze}, {Bonnet-Bidaud}, {Mouchet}, {Coppejans}, and
  {Gulbis}]{2014MNRAS.444.3004D}
{de Martino}, D.; {Casares}, J.; {Mason}, E.; {Buckley}, D.A.H.; {Kotze}, M.M.;
  {Bonnet-Bidaud}, J.M.; {Mouchet}, M.; {Coppejans}, R.; {Gulbis}, A.A.S.
\newblock {Unveiling the redback nature of the low-mass X-ray binary XSS
  J1227.0-4859 through optical observations}.
\newblock {\em Mon. Not. R. Astron. Soc.} {\bf 2014}, {\em 444},~3004--3014,
\newblock
  doi:{\changeurlcolor{black}\href{https://doi.org/10.1093/mnras/stu1640}{\detokenize{10.1093/mnras/stu1640}}}.

\bibitem[{Xing} and {Wang}(2015)]{2015ApJ...808...17X}
{Xing}, Y.; {Wang}, Z.
\newblock {Fermi Observation of the Transitional Pulsar Binary XSS
  J12270-4859}.
\newblock {\em  Astrophys. J.} {\bf 2015}, {\em 808},~17,
\newblock
  doi:{\changeurlcolor{black}\href{https://doi.org/10.1088/0004-637X/808/1/17}{\detokenize{10.1088/0004-637X/808/1/17}}}.

\bibitem[]{2015MNRAS.449L..26P} Papitto, A.; de Martino, D.; Belloni, T.M.; Burgay, M.; Pellizzoni, A.; Possenti, A.; Torres, D.F. X-ray coherent pulsations during a sub-luminous accretion disc state of the transitional millisecond pulsar XSS J12270-4859. {\em Mon. Not. R.  Astron. Soc.} {\bf 2015}, {\em 449}, 26.

\bibitem[{Ray} \em{et~al.}(2012){Ray}, {Abdo}, {Parent}, {Bhattacharya},
  {Bhattacharyya}, {Camilo}, {Cognard}, {Theureau}, {Ferrara}, {Harding},
  {Thompson}, {Freire}, {Guillemot}, {Gupta}, {Roy}, {Hessels}, {Johnston},
  {Keith}, {Shannon}, {Kerr}, {Michelson}, {Romani}, {Kramer}, {McLaughlin},
  {Ransom}, {Roberts}, {Saz Parkinson}, {Ziegler}, {Smith}, {Stappers},
  {Weltevrede}, and {Wood}]{2012arXiv1205.3089R}
{Ray}, P.S.; {Abdo}, A.A.; {Parent}, D.; {Bhattacharya}, D.; {Bhattacharyya},
  B.; {Camilo}, F.; {Cognard}, I.; {Theureau},~G.; {Ferrara}, E.C.; {Harding},
  A.K.; et al.
\newblock {Radio Searches of Fermi LAT Sources and Blind Search Pulsars: The~Fermi Pulsar Search Consortium}.
\newblock \emph{arXiv} {\bf 2012}, arXiv:1205.3089.

\bibitem[{Li} \em{et~al.}(2014){Li}, {Halpern}, and
  {Thorstensen}]{2014ApJ...795..115L}
{Li}, M.; {Halpern}, J.P.; {Thorstensen}, J.R.
\newblock {Optical Counterparts of Two Fermi Millisecond Pulsars: PSR~J1301+0833 and PSR J1628-3205}.
\newblock {\em  Astrophys. J.} {\bf 2014}, {\em 795},~115,
\newblock
  doi:{\changeurlcolor{black}\href{https://doi.org/10.1088/0004-637X/795/2/115}{\detokenize{10.1088/0004-637X/795/2/115}}}.

\bibitem[{Pletsch} \em{et~al.}(2012){Pletsch}, {Guillemot}, {Allen}, {Kramer},
  {Aulbert}, {Fehrmann}, {Ray}, {Barr}, {Belfiore}, {Camilo}, {Caraveo},
  {{\c{C}}elik}, {Champion}, {Dormody}, {Eatough}, {Ferrara}, {Freire},
  {Hessels}, {Keith}, {Kerr}, {de Luca}, {Lyne}, {Marelli}, {McLaughlin},
  {Parent}, {Ransom}, {Razzano}, {Reich}, {Saz Parkinson}, {Stappers}, and
  {Wolff}]{2012ApJ...744..105P}
{Pletsch}, H.J.; {Guillemot}, L.; {Allen}, B.; {Kramer}, M.; {Aulbert}, C.;
  {Fehrmann}, H.; {Ray}, P.S.; {Barr}, E.D.; {Belfiore},~A.; {Camilo}, F.;
  et al.
\newblock {Discovery of Nine Gamma-Ray Pulsars in Fermi Large Area Telescope
  Data Using a New Blind Search Method}.
\newblock {\em  Astrophys. J.} {\bf 2012}, {\em 744},~105,
\newblock
  doi:{\changeurlcolor{black}\href{https://doi.org/10.1088/0004-637X/744/2/105}{\detokenize{10.1088/0004-637X/744/2/105}}}.

\bibitem[{Romani} \em{et~al.}(2012){Romani}, {Filippenko}, {Silverman},
  {Cenko}, {Greiner}, {Rau}, {Elliott}, and {Pletsch}]{2012ApJ...760L..36R}
{Romani}, R.W.; {Filippenko}, A.V.; {Silverman}, J.M.; {Cenko}, S.B.;
  {Greiner}, J.; {Rau}, A.; {Elliott}, J.; {Pletsch}, H.J.
\newblock {PSR~J1311-3430: A Heavyweight Neutron Star with a Flyweight Helium
  Companion}.
\newblock {\em Astrophys. J. Lett.} {\bf 2012}, {\em 760},~L36,
\newblock
  doi:{\changeurlcolor{black}\href{https://doi.org/10.1088/2041-8205/760/2/L36}{\detokenize{10.1088/2041-8205/760/2/L36}}}.

\bibitem[{Xing} and {Wang}(2015)]{2015ApJ...804L..33X}
{Xing}, Y.; {Wang}, Z.
\newblock {Discovery of Gamma-Ray Orbital Modulation in the Black Widow PSR
  J1311-3430}.
\newblock {\em Astrophys. J. Lett.} {\bf 2015}, {\em 804},~L33,
\newblock
  doi:{\changeurlcolor{black}\href{https://doi.org/10.1088/2041-8205/804/2/L33}{\detokenize{10.1088/2041-8205/804/2/L33}}}.

\bibitem[{Bates} \em{et~al.}(2015){Bates}, {Thornton}, {Bailes}, {Barr},
  {Bassa}, {Bhat}, {Burgay}, {Burke-Spolaor}, {Champion}, {Flynn}, {Jameson},
  {Johnston}, {Keith}, {Kramer}, {Levin}, {Lyne}, {Milia}, {Ng}, {Petroff},
  {Possenti}, {Stappers}, {van Straten}, and {Tiburzi}]{2015MNRAS.446.4019B}
{Bates}, S.D.; {Thornton}, D.; {Bailes}, M.; {Barr}, E.; {Bassa}, C.G.; {Bhat},
  N.D.R.; {Burgay}, M.; {Burke-Spolaor}, S.; {Champion}, D.J.; {Flynn}, C.M.L.;
  et al.
\newblock {The High Time Resolution Universe survey - XI. Discovery of five
  recycled pulsars and the optical detectability of survey white dwarf
  companions}.
\newblock {\em Mon. Not. R. Astron. Soc.} {\bf 2015}, {\em 446},~4019--4028,
\newblock
  doi:{\changeurlcolor{black}\href{https://doi.org/10.1093/mnras/stu2350}{\detokenize{10.1093/mnras/stu2350}}}.

\bibitem[{Strader} \em{et~al.}(2019){Strader}, {Swihart}, {Chomiuk},
  {Bahramian}, {Britt}, {Cheung}, {Dage}, {Halpern}, {Li}, {Mignani}, {Orosz},
  {Peacock}, {Salinas}, {Shishkovsky}, and {Tremou}]{2019ApJ...872...42S}
{Strader}, J.; {Swihart}, S.; {Chomiuk}, L.; {Bahramian}, A.; {Britt}, C.;
  {Cheung}, C.C.; {Dage}, K.; {Halpern}, J.; {Li}, K.L.; {Mignani}, R.P.;
  et al.
\newblock {Optical Spectroscopy and Demographics of Redback Millisecond Pulsar
  Binaries}.
\newblock {\em  Astrophys. J.} {\bf 2019}, {\em 872},~42,
\newblock
  doi:{\changeurlcolor{black}\href{https://doi.org/10.3847/1538-4357/aafbaa}{\detokenize{10.3847/1538-4357/aafbaa}}}.

\bibitem[{Ng} \em{et~al.}(2014){Ng}, {Bailes}, {Bates}, {Bhat}, {Burgay},
  {Burke-Spolaor}, {Champion}, {Coster}, {Johnston}, {Keith}, {Kramer},
  {Levin}, {Petroff}, {Possenti}, {Stappers}, {van Straten}, {Thornton},
  {Tiburzi}, {Bassa}, {Freire}, {Guillemot}, {Lyne}, {Tauris}, {Shannon}, and
  {Wex}]{2014MNRAS.439.1865N}
{Ng}, C.; {Bailes}, M.; {Bates}, S.D.; {Bhat}, N.D.R.; {Burgay}, M.;
  {Burke-Spolaor}, S.; {Champion}, D.J.; {Coster}, P.; {Johnston}, S.; {Keith},
  M.J.; et al.
\newblock {The High Time Resolution Universe pulsar survey---X. Discovery of
  four millisecond pulsars and updated timing solutions of a further 12}.
\newblock {\em Mon. Not. R. Astron. Soc.} {\bf 2014}, {\em 439},~1865--1883,
\newblock
  doi:{\changeurlcolor{black}\href{https://doi.org/10.1093/mnras/stu067}{\detokenize{10.1093/mnras/stu067}}}.

\bibitem[{Sanpa-arsa}(2016)]{san16}
{Sanpa-arsa}, S.
\newblock Searching for New Millisecond Pulsars with the Gbt in Fermi
  Unassociated Sources.
\newblock Ph.D~Thesis, University of Virginia,  Charlottesville, VA, USA, 2016.

\bibitem[{Bhattacharyya} \em{et~al.}(2013){Bhattacharyya}, {Roy}, {Ray},
  {Gupta}, {Bhattacharya}, {Romani}, {Ransom}, {Ferrara}, {Wolff}, {Camilo},
  {Cognard}, {Harding}, {den Hartog}, {Johnston}, {Keith}, {Kerr}, {Michelson},
  {Saz Parkinson}, {Wood}, and {Wood}]{2013ApJ...773L..12B}
{Bhattacharyya}, B.; {Roy}, J.; {Ray}, P.S.; {Gupta}, Y.; {Bhattacharya}, D.;
  {Romani}, R.W.; {Ransom}, S.M.; {Ferrara},~E.C.; {Wolff}, M.T.; {Camilo}, F.;
  et al.
\newblock {GMRT Discovery of PSR J1544+4937: An Eclipsing Black-widow Pulsar
  Identified with a Fermi-LAT Source}.
\newblock {\em Astrophys. J. Lett.} {\bf 2013}, {\em 773},~L12,
\newblock
  doi:{\changeurlcolor{black}\href{https://doi.org/10.1088/2041-8205/773/1/L12}{\detokenize{10.1088/2041-8205/773/1/L12}}}.

\bibitem[{Tang} \em{et~al.}(2014){Tang}, {Kaplan}, {Phinney}, {Prince},
  {Breton}, {Bellm}, {Bildsten}, {Cao}, {Kong}, {Perley}, {Sesar}, {Wolf}, and
  {Yen}]{2014ApJ...791L...5T}
{Tang}, S.; {Kaplan}, D.L.; {Phinney}, E.S.; {Prince}, T.A.; {Breton}, R.P.;
  {Bellm}, E.; {Bildsten}, L.; {Cao}, Y.; {Kong}, A.K.H.; {Perley}, D.A.;
  et al.
\newblock {Identification of the Optical Counterpart of Fermi Black Widow
  Millisecond Pulsar PSR J1544+4937}.
\newblock {\em Astrophys. J. Lett.} {\bf 2014}, {\em 791},~L5,
\newblock
  doi:{\changeurlcolor{black}\href{https://doi.org/10.1088/2041-8205/791/1/L5}{\detokenize{10.1088/2041-8205/791/1/L5}}}.

\bibitem[{Lorimer}(2019)]{wvu}
{Lorimer}, D.
\newblock All Published and Unpublished Millisecond Pulsars Not Associated with
  a Globular Cluster.
\newblock Available online: \url{http://astro.phys.wvu.edu/GalacticMSPs/GalacticMSPs.txt}
  (accessed on 29 July 2019).

\bibitem[{Lynch} \em{et~al.}(2018){Lynch}, {Swiggum}, {Kondratiev}, {Kaplan},
  {Stovall}, {Fonseca}, {Roberts}, {Levin}, {DeCesar}, {Cui}, {Cenko},
  {Gatkine}, {Archibald}, {Banaszak}, {Biwer}, {Boyles}, {Chawla}, {Dartez},
  {Day}, {Ford}, {Flanigan}, {Hessels}, {Hinojosa}, {Jenet}, {Karako-Argaman},
  {Kaspi}, {Leake}, {Lunsford}, {Martinez}, {Mata}, {McLaughlin}, {Noori},
  {Ransom}, {Rohr}, {Siemens}, {Spiewak}, {Stairs}, {van Leeuwen}, {Walker},
  and {Wells}]{2018ApJ...859...93L}
{Lynch}, R.S.; {Swiggum}, J.K.; {Kondratiev}, V.I.; {Kaplan}, D.L.; {Stovall},
  K.; {Fonseca}, E.; {Roberts}, M.S.E.; {Levin}, L.; {DeCesar}, M.E.; {Cui},
  B.; et al.
\newblock {The Green Bank North Celestial Cap Pulsar Survey. III. 45 New Pulsar
  Timing Solutions}.
\newblock {\em  Astrophys. J.} {\bf 2018}, {\em 859},~93,
\newblock
  doi:{\changeurlcolor{black}\href{https://doi.org/10.3847/1538-4357/aabf8a}{\detokenize{10.3847/1538-4357/aabf8a}}}.

\bibitem[{Crawford} \em{et~al.}(2013){Crawford}, {Lyne}, {Stairs}, {Kaplan},
  {McLaughlin}, {Freire}, {Burgay}, {Camilo}, {D'Amico}, {Faulkner}, {Kramer},
  {Lorimer}, {Manchester}, {Possenti}, and {Steeghs}]{2013ApJ...776...20C}
{Crawford}, F.; {Lyne}, A.G.; {Stairs}, I.H.; {Kaplan}, D.L.; {McLaughlin},
  M.A.; {Freire}, P.C.C.; {Burgay}, M.; {Camilo}, F.; {D'Amico}, N.;
  {Faulkner}, A.; et al.
\newblock {PSR J1723-2837: An Eclipsing Binary Radio Millisecond Pulsar}.
\newblock {\em  Astrophys.~J.} {\bf 2013}, {\em 776},~20,
\newblock
  doi:{\changeurlcolor{black}\href{https://doi.org/10.1088/0004-637X/776/1/20}{\detokenize{10.1088/0004-637X/776/1/20}}}.

\bibitem[{Kong} \em{et~al.}(2017){Kong}, {Hui}, {Takata}, {Li}, and
  {Tam}]{2017ApJ...839..130K}
{Kong}, A.K.H.; {Hui}, C.Y.; {Takata}, J.; {Li}, K.L.; {Tam}, P.H.T.
\newblock {A NuSTAR Observation of the Gamma-Ray Emitting Millisecond Pulsar
  PSR J1723-2837}.
\newblock {\em  Astrophys. J.} {\bf 2017}, {\em 839},~130,
\newblock
  doi:{\changeurlcolor{black}\href{https://doi.org/10.3847/1538-4357/aa6aa2}{\detokenize{10.3847/1538-4357/aa6aa2}}}.

\bibitem[{Barr} \em{et~al.}(2013){Barr}, {Guillemot}, {Champion}, {Kramer},
  {Eatough}, {Lee}, {Verbiest}, {Bassa}, {Camilo}, {{\c{C}}elik}, {Cognard},
  {Ferrara}, {Freire}, {Janssen}, {Johnston}, {Keith}, {Lyne}, {Michelson},
  {Parkinson}, {Ransom}, {Ray}, {Stappers}, and {Wood}]{2013MNRAS.429.1633B}
{Barr}, E.D.; {Guillemot}, L.; {Champion}, D.J.; {Kramer}, M.; {Eatough}, R.P.;
  {Lee}, K.J.; {Verbiest}, J.P.W.; {Bassa}, C.G.; {Camilo}, F.; {{\c{C}}elik},
  {\"O}.; et al.
\newblock {Pulsar searches of Fermi unassociated sources with the Effelsberg
  telescope}.
\newblock {\em Mon.~Not. R. Astron. Soc.} {\bf 2013}, {\em 429},~1633--1642,
\newblock
  doi:{\changeurlcolor{black}\href{https://doi.org/10.1093/mnras/sts449}{\detokenize{10.1093/mnras/sts449}}}.

\bibitem[{Stovall} \em{et~al.}(2014){Stovall}, {Lynch}, {Ransom}, {Archibald},
  {Banaszak}, {Biwer}, {Boyles}, {Dartez}, {Day}, {Ford}, {Flanigan}, {Garcia},
  {Hessels}, {Hinojosa}, {Jenet}, {Kaplan}, {Karako-Argaman}, {Kaspi},
  {Kondratiev}, {Leake}, {Lorimer}, {Lunsford}, {Martinez}, {Mata},
  {McLaughlin}, {Roberts}, {Rohr}, {Siemens}, {Stairs}, {van Leeuwen},
  {Walker}, and {Wells}]{2014ApJ...791...67S}
{Stovall}, K.; {Lynch}, R.S.; {Ransom}, S.M.; {Archibald}, A.M.; {Banaszak},
  S.; {Biwer}, C.M.; {Boyles}, J.; {Dartez}, L.P.; {Day}, D.; {Ford}, A.J.;
  et al.
\newblock {The Green Bank Northern Celestial Cap Pulsar Survey. I. Survey
  Description, Data Analysis, and Initial Results}.
\newblock {\em  Astrophys. J.} {\bf 2014}, {\em 791},~67,
\newblock
  doi:{\changeurlcolor{black}\href{https://doi.org/10.1088/0004-637X/791/1/67}{\detokenize{10.1088/0004-637X/791/1/67}}}.

\bibitem[{Kaplan} \em{et~al.}(2012){Kaplan}, {Stovall}, {Ransom}, {Roberts},
  {Kotulla}, {Archibald}, {Biwer}, {Boyles}, {Dartez}, {Day}, {Ford}, {Garcia},
  {Hessels}, {Jenet}, {Karako}, {Kaspi}, {Kondratiev}, {Lorimer}, {Lynch},
  {McLaughlin}, {Rohr}, {Siemens}, {Stairs}, and {van
  Leeuwen}]{2012ApJ...753..174K}
{Kaplan}, D.L.; {Stovall}, K.; {Ransom}, S.M.; {Roberts}, M.S.E.; {Kotulla},
  R.; {Archibald}, A.M.; {Biwer}, C.M.; {Boyles}, J.; {Dartez}, L.; {Day},
  D.F.; et al.
\newblock {Discovery of the Optical/Ultraviolet/Gamma-Ray Counterpart to the
  Eclipsing Millisecond Pulsar J1816+4510}.
\newblock {\em  Astrophys. J.} {\bf 2012}, {\em 753},~174,
\newblock
  doi:{\changeurlcolor{black}\href{https://doi.org/10.1088/0004-637X/753/2/174}{\detokenize{10.1088/0004-637X/753/2/174}}}.

\bibitem[Parent et al.(2019)]{2019arXiv190809926P} Parent, E.; Kaspi, V.M.; Ransom, S.M.; Freire, P.C.C.; Brazier, A.; Camilo, F.; Chatterjee, S.; Cordes, J.M.; Crawford, F.; Deneva, J.S.; et al. Eight Millisecond Pulsars Discovered in the Arecibo PALFA Survey. {\em  Astrophys. J.} {\bf 2019}, {\em 886},~148.

\bibitem[{Stovall} \em{et~al.}(2016){Stovall}, {Allen}, {Bogdanov}, {Brazier},
  {Camilo}, {Cardoso}, {Chatterjee}, {Cordes}, {Crawford}, {Deneva}, {Ferdman},
  {Freire}, {Hessels}, {Jenet}, {Kaplan}, {Karako-Argaman}, {Kaspi}, {Knispel},
  {Kotulla}, {Lazarus}, {Lee}, {van Leeuwen}, {Lynch}, {Lyne}, {Madsen},
  {McLaughlin}, {Patel}, {Ransom}, {Scholz}, {Siemens}, {Stairs}, {Stappers},
  {Swiggum}, {Zhu}, and {Venkataraman}]{2016ApJ...833..192S}
{Stovall}, K.; {Allen}, B.; {Bogdanov}, S.; {Brazier}, A.; {Camilo}, F.;
  {Cardoso}, F.; {Chatterjee}, S.; {Cordes}, J.M.; {Crawford},~F.; {Deneva},
  J.S.; et al.
\newblock {Timing of Five PALFA-discovered Millisecond Pulsars}.
\newblock {\em  Astrophys. J.} {\bf 2016}, {\em 833},~192,
\newblock
  doi:{\changeurlcolor{black}\href{https://doi.org/10.3847/1538-4357/833/2/192}{\detokenize{10.3847/1538-4357/833/2/192}}}.

\bibitem[]{fruchter1988}
Fruchter, A.S.; Stinebring, D.R.; Taylor, J.H. A millisecond pulsar in an eclipsing binary. {\em Nature} {\bf 1988}, {\em 333}, 237--239.

\bibitem[{Reynolds} \em{et~al.}(2007){Reynolds}, {Callanan}, {Fruchter},
  {Torres}, {Beer}, and {Gibbons}]{2007MNRAS.379.1117R}
{Reynolds}, M.T.; {Callanan}, P.J.; {Fruchter}, A.S.; {Torres}, M.A.P.; {Beer},
  M.E.; {Gibbons}, R.A.
\newblock {The light curve of the companion to PSR B1957+20}.
\newblock {\em Mon. Not. R. Astron. Soc.} {\bf 2007}, {\em 379},~1117--1122,
\newblock
  doi:{\changeurlcolor{black}\href{https://doi.org/10.1111/j.1365-2966.2007.11991.x}{\detokenize{10.1111/j.1365-2966.2007.11991.x}}}.

\bibitem[Huang et al. (2012)]{huang2012}
Huang, R.H.H.; Kong, A.K.H.; Takata, J.; Hui, C.Y.; Lin, L.C.C.; Cheng, K.S. X-ray Studies of the Black Widow Pulsar PSR B1957+20. {\em \apj} {\bf 2012}, {\em 760}, 92.

\bibitem[Wu et al. (2012)]{wu2012}
Wu, E.M.H.; Takata, J.; Cheng, K.S.; Huang, R.H.H.; Hui, C.Y.; Kong, A.K.H.; Tam, P.H.T.; Wu, J.H.K. Orbital-phase-dependent Gamma-Ray Emissions from the Black Widow Pulsar. {\em \apj} {\bf 2012}, {\em 761}, 181.

\bibitem[{Shaifullah} \em{et~al.}(2016){Shaifullah}, {Verbiest}, {Freire},
  {Tauris}, {Wex}, {Os{\l}owski}, {Stappers}, {Bassa}, {Caballero}, {Champion},
  {Cognard}, {Desvignes}, {Graikou}, {Guillemot}, {Janssen}, {Jessner},
  {Jordan}, {Karuppusamy}, {Kramer}, {Lazaridis}, {Lazarus}, {Lyne}, {McKee},
  {Perrodin}, {Possenti}, and {Tiburzi}]{2016MNRAS.462.1029S}
{Shaifullah}, G.; {Verbiest}, J.P.W.; {Freire}, P.C.C.; {Tauris}, T.M.; {Wex},
  N.; {Os{\l}owski}, S.; {Stappers}, B.W.; {Bassa}, C.G.; {Caballero}, R.N.;
  {Champion}, D.J.; et al.
\newblock {21 year timing of the black-widow pulsar J2051-0827}.
\newblock {\em Mon. Not. R. Astron. Soc.} {\bf 2016}, {\em 462},~1029--1038,
\newblock
  doi:{\changeurlcolor{black}\href{https://doi.org/10.1093/mnras/stw1737}{\detokenize{10.1093/mnras/stw1737}}}.

\bibitem[{Stappers} \em{et~al.}(1999){Stappers}, {van Kerkwijk}, {Lane}, and
  {Kulkarni}]{1999ApJ...510L..45S}
{Stappers}, B.W.; {van Kerkwijk}, M.H.; {Lane}, B.; {Kulkarni}, S.R.
\newblock {The Light Curve of the Companion to PSR J2051-0827}.
\newblock {\em Astrophys. J. Lett.} {\bf 1999}, {\em 510},~L45--L48,
\newblock
  doi:{\changeurlcolor{black}\href{https://doi.org/10.1086/311795}{\detokenize{10.1086/311795}}}.

\bibitem[{Guillemot} \em{et~al.}(2019){Guillemot}, {Octau}, {Cognard},
  {Desvignes}, {Freire}, {Smith}, {Theureau}, and
  {Burnett}]{2019arXiv190709778G}
{Guillemot}, L.; {Octau}, F.; {Cognard}, I.; {Desvignes}, G.; {Freire}, P.C.C.;
  {Smith}, D.A.; {Theureau}, G.; {Burnett}, T.H.
\newblock {Timing of PSR J2055+3829, an eclipsing black widow pulsar discovered
  with the Nançay Radio Telescope}.
\newblock {\em arXiv} {\bf 2019}, arXiv:1907.09778.


\bibitem[{Bellm} \em{et~al.}(2016){Bellm}, {Kaplan}, {Breton}, {Phinney},
  {Bhalerao}, {Camilo}, {Dahal}, {Djorgovski}, {Drake}, {Hessels}, {Laher},
  {Levitan}, {Lewis}, {Mahabal}, {Ofek}, {Prince}, {Ransom}, {Roberts},
  {Russell}, {Sesar}, {Surace}, and {Tang}]{2016ApJ...816...74B}
{Bellm}, E.C.; {Kaplan}, D.L.; {Breton}, R.P.; {Phinney}, E.S.; {Bhalerao},
  V.B.; {Camilo}, F.; {Dahal}, S.; {Djorgovski}, S.G.; {Drake}, A.J.;
  {Hessels}, J.W.T.; et al.
\newblock {Properties and Evolution of the Redback Millisecond Pulsar Binary
  PSR J2129-0429}.
\newblock {\em  Astrophys. J.} {\bf 2016}, {\em 816},~74,
\newblock
  doi:{\changeurlcolor{black}\href{https://doi.org/10.3847/0004-637X/816/2/74}{\detokenize{10.3847/0004-637X/816/2/74}}}.

\bibitem[{Kong} \em{et~al.}(2018){Kong}, {Takata}, {Hui}, {Zhao}, {Li}, and
  {Tam}]{2018MNRAS.478.3987K}
{Kong}, A.K.H.; {Takata}, J.; {Hui}, C.Y.; {Zhao}, J.; {Li}, K.L.; {Tam},
  P.H.T.
\newblock {Broad-band high-energy emissions of the redback millisecond pulsar
  PSR J2129-0429}.
\newblock {\em Mon. Not. R. Astron. Soc.} {\bf 2018}, {\em 478},~3987--3993,
\newblock
  doi:{\changeurlcolor{black}\href{https://doi.org/10.1093/mnras/sty1459}{\detokenize{10.1093/mnras/sty1459}}}.

\bibitem[{Schroeder} and {Halpern}(2014)]{2014ApJ...793...78S}
{Schroeder}, J.; {Halpern}, J.
\newblock {Observations and Modeling of the Companions of Short Period Binary
  Millisecond Pulsars: Evidence for High-mass Neutron Stars}.
\newblock {\em  Astrophys. J.} {\bf 2014}, {\em 793},~78,
\newblock
  doi:{\changeurlcolor{black}\href{https://doi.org/10.1088/0004-637X/793/2/78}{\detokenize{10.1088/0004-637X/793/2/78}}}.

\bibitem[{Keith} \em{et~al.}(2011){Keith}, {Johnston}, {Ray}, {Ferrara}, {Saz
  Parkinson}, {{\c{C}}elik}, {Belfiore}, {Donato}, {Cheung}, {Abdo}, {Camilo},
  {Freire}, {Guillemot}, {Harding}, {Kramer}, {Michelson}, {Ransom}, {Romani},
  {Smith}, {Thompson}, {Weltevrede}, and {Wood}]{2011MNRAS.414.1292K}
{Keith}, M.J.; {Johnston}, S.; {Ray}, P.S.; {Ferrara}, E.C.; {Saz Parkinson},
  P.M.; {{\c{C}}elik}, {\"O}.; {Belfiore}, A.; {Donato}, D.; {Cheung}, C.C.;
  {Abdo}, A.A.; et al.
\newblock {Discovery of millisecond pulsars in radio searches of southern Fermi
  Large Area Telescope sources}.
\newblock {\em Mon. Not. R. Astron. Soc.} {\bf 2011}, {\em 414},~1292--1300,
\newblock
  doi:{\changeurlcolor{black}\href{https://doi.org/10.1111/j.1365-2966.2011.18464.x}{\detokenize{10.1111/j.1365-2966.2011.18464.x}}}.

\bibitem[{An} \em{et~al.}(2018){An}, {Romani}, and {Kerr}]{2018ApJ...868L...8A}
{An}, H.; {Romani}, R.W.; {Kerr}, M.
\newblock {Signatures of Intra-binary Shock Emission in the Black Widow Pulsar
  Binary PSR J2241-5236}.
\newblock {\em Astrophys. J. Lett.} {\bf 2018}, {\em 868},~L8,
\newblock
  doi:{\changeurlcolor{black}\href{https://doi.org/10.3847/2041-8213/aaedaf}{\detokenize{10.3847/2041-8213/aaedaf}}}.

\bibitem[{Pletsch} and {Clark}(2015)]{2015ApJ...807...18P}
{Pletsch}, H.J.; {Clark}, C.J.
\newblock {Gamma-Ray Timing of Redback PSR J2339-0533: Hints for Gravitational
  Quadrupole Moment Changes}.
\newblock {\em  Astrophys. J.} {\bf 2015}, {\em 807},~18,
\newblock
  doi:{\changeurlcolor{black}\href{https://doi.org/10.1088/0004-637X/807/1/18}{\detokenize{10.1088/0004-637X/807/1/18}}}.


\bibitem[{Camilo} \em{et~al.}(2000){Camilo}, {Lorimer}, {Freire}, {Lyne}, and
  {Manchester}]{2000ApJ...535..975C}
{Camilo}, F.; {Lorimer}, D.R.; {Freire}, P.; {Lyne}, A.G.; {Manchester}, R.N.
\newblock {Observations of 20 Millisecond Pulsars in 47 Tucanae at 20
  Centimeters}.
\newblock {\em  Astrophys. J.} {\bf 2000}, {\em 535},~975--990,
\newblock
  doi:{\changeurlcolor{black}\href{https://doi.org/10.1086/308859}{\detokenize{10.1086/308859}}}.

\bibitem[{Becker} \em{et~al.}(2010){Becker}, {Huang}, and
  {Prinz}]{2010arXiv1006.0335B}
{Becker}, W.; {Huang}, H.H.; {Prinz}, T.
\newblock {X-ray Counterparts of Millisecond Pulsars in Globular Clusters}.
\newblock {\em arXiv}~{\bf 2010},~arXiv:1006.0335.

\bibitem[{Cadelano} \em{et~al.}(2015){Cadelano}, {Pallanca}, {Ferraro},
  {Salaris}, {Dalessandro}, {Lanzoni}, and {Freire}]{2015ApJ...812...63C}
{Cadelano}, M.; {Pallanca}, C.; {Ferraro}, F.R.; {Salaris}, M.; {Dalessandro},
  E.; {Lanzoni}, B.; {Freire}, P.C.C.
\newblock {Optical~Identification of He White Dwarfs Orbiting Four Millisecond
  Pulsars in the Globular Cluster 47 Tucanae}.
\newblock {\em  Astrophys. J.} {\bf 2015}, {\em 812},~63,
\newblock
  doi:{\changeurlcolor{black}\href{https://doi.org/10.1088/0004-637X/812/1/63}{\detokenize{10.1088/0004-637X/812/1/63}}}.

\bibitem[{Bogdanov} \em{et~al.}(2005){Bogdanov}, {Grindlay}, and {van den
  Berg}]{2005ApJ...630.1029B}
{Bogdanov}, S.; {Grindlay}, J.E.; {van den Berg}, M.
\newblock {An X-ray Variable Millisecond Pulsar in the Globular Cluster 47
  Tucanae: Closing the Link to Low-Mass X-ray Binaries}.
\newblock {\em  Astrophys. J.} {\bf 2005}, {\em 630},~1029--1036,
\newblock
  doi:{\changeurlcolor{black}\href{https://doi.org/10.1086/432249}{\detokenize{10.1086/432249}}}.

\bibitem[{Pallanca} \em{et~al.}(2014){Pallanca}, {Ransom}, {Ferraro},
  {Dalessand ro}, {Lanzoni}, {Hessels}, {Stairs}, and
  {Freire}]{2014ApJ...795...29P}
{Pallanca}, C.; {Ransom}, S.M.; {Ferraro}, F.R.; {Dalessand ro}, E.; {Lanzoni},
  B.; {Hessels}, J.W.T.; {Stairs}, I.; {Freire}, P.C.C.
\newblock {Radio Timing and Optical Photometry of the Black Widow System PSR
  J1518+0204C in the Globular Cluster M5}.
\newblock {\em  Astrophys. J.} {\bf 2014}, {\em 795},~29,
\newblock
  doi:{\changeurlcolor{black}\href{https://doi.org/10.1088/0004-637X/795/1/29}{\detokenize{10.1088/0004-637X/795/1/29}}}.

\bibitem[{Hessels} \em{et~al.}(2007){Hessels}, {Ransom}, {Stairs}, {Kaspi}, and
  {Freire}]{2007ApJ...670..363H}
{Hessels}, J.W.T.; {Ransom}, S.M.; {Stairs}, I.H.; {Kaspi}, V.M.; {Freire},
  P.C.C.
\newblock {A 1.4 GHz Arecibo Survey for Pulsars in Globular Clusters}.
\newblock {\em  Astrophys. J.} {\bf 2007}, {\em 670},~363--378,
\newblock
  doi:{\changeurlcolor{black}\href{https://doi.org/10.1086/521780}{\detokenize{10.1086/521780}}}.

\bibitem[{Lynch} \em{et~al.}(2012){Lynch}, {Freire}, {Ransom}, and
  {Jacoby}]{2012ApJ...745..109L}
{Lynch}, R.S.; {Freire}, P.C.C.; {Ransom}, S.M.; {Jacoby}, B.A.
\newblock {The Timing of Nine Globular Cluster Pulsars}.
\newblock {\em  Astrophys. J.} {\bf 2012}, {\em 745},~109,
\newblock
  doi:{\changeurlcolor{black}\href{https://doi.org/10.1088/0004-637X/745/2/109}{\detokenize{10.1088/0004-637X/745/2/109}}}.

\bibitem[{Cocozza} \em{et~al.}(2008){Cocozza}, {Ferraro}, {Possenti},
  {Beccari}, {Lanzoni}, {Ransom}, {Rood}, and {D'Amico}]{2008ApJ...679L.105C}
{Cocozza}, G.; {Ferraro}, F.R.; {Possenti}, A.; {Beccari}, G.; {Lanzoni}, B.;
  {Ransom}, S.; {Rood}, R.T.; {D'Amico},~N.
\newblock {A~Puzzling Millisecond Pulsar Companion in NGC 6266}.
\newblock {\em Astrophys. J. Lett.} {\bf 2008}, {\em 679},~L105,
\newblock
  doi:{\changeurlcolor{black}\href{https://doi.org/10.1086/589557}{\detokenize{10.1086/589557}}}.

\bibitem[{D'Amico} \em{et~al.}(2001){D'Amico}, {Possenti}, {Manchester},
  {Sarkissian}, {Lyne}, and {Camilo}]{2001ApJ...561L..89D}
{D'Amico}, N.; {Possenti}, A.; {Manchester}, R.N.; {Sarkissian}, J.; {Lyne},
  A.G.; {Camilo}, F.
\newblock {An Eclipsing Millisecond Pulsar with a Possible Main-Sequence
  Companion in NGC 6397}.
\newblock {\em Astrophys. J. Lett.} {\bf 2001}, {\em 561},~L89--L92,
\newblock
  doi:{\changeurlcolor{black}\href{https://doi.org/10.1086/324562}{\detokenize{10.1086/324562}}}.

\bibitem[{Kaluzny} \em{et~al.}(2003){Kaluzny}, {Rucinski}, and
  {Thompson}]{2003AJ....125.1546K}
{Kaluzny}, J.; {Rucinski}, S.M.; {Thompson}, I.B.
\newblock {Photometry and Spectroscopy of the Optical Companion to the Pulsar
  PSR J1740-5340 in the Globular Cluster NGC 6397}.
\newblock {\em \aj} {\bf 2003}, {\em 125},~1546--1553,
\newblock
  doi:{\changeurlcolor{black}\href{https://doi.org/10.1086/346273}{\detokenize{10.1086/346273}}}.

\bibitem[{Freire} \em{et~al.}(2008){Freire}, {Ransom}, {B{\'e}gin}, {Stairs},
  {Hessels}, {Frey}, and {Camilo}]{2008ApJ...675..670F}
{Freire}, P.C.C.; {Ransom}, S.M.; {B{\'e}gin}, S.; {Stairs}, I.H.; {Hessels},
  J.W.T.; {Frey}, L.H.; {Camilo}, F.
\newblock {Eight New Millisecond Pulsars in NGC 6440 and NGC 6441}.
\newblock {\em  Astrophys. J.} {\bf 2008}, {\em 675},~670--682,
\newblock
  doi:{\changeurlcolor{black}\href{https://doi.org/10.1086/526338}{\detokenize{10.1086/526338}}}.

\bibitem[{Hobbs} \em{et~al.}(2004){Hobbs}, {Faulkner}, {Stairs}, {Camilo},
  {Manchester}, {Lyne}, {Kramer}, {D'Amico}, {Kaspi}, {Possenti}, {McLaughlin},
  {Lorimer}, {Burgay}, {Joshi}, and {Crawford}]{2004MNRAS.352.1439H}
{Hobbs}, G.; {Faulkner}, A.; {Stairs}, I.H.; {Camilo}, F.; {Manchester}, R.N.;
  {Lyne}, A.G.; {Kramer}, M.; {D'Amico}, N.; {Kaspi}, V.M.; {Possenti}, A.;
  et al.
\newblock {The Parkes multibeam pulsar survey---IV. Discovery of 180 pulsars
  and parameters for 281 previously known pulsars}.
\newblock {\em Mon. Not. R. Astron. Soc.} {\bf 2004}, {\em 352},~1439--1472,
\newblock
  doi:{\changeurlcolor{black}\href{https://doi.org/10.1111/j.1365-2966.2004.08042.x}{\detokenize{10.1111/j.1365-2966.2004.08042.x}}}.

\bibitem[{Ransom} \em{et~al.}(2005){Ransom}, {Hessels}, {Stairs}, {Freire},
  {Camilo}, {Kaspi}, and {Kaplan}]{2005Sci...307..892R}
{Ransom}, S.M.; {Hessels}, J.W.T.; {Stairs}, I.H.; {Freire}, P.C.C.; {Camilo},
  F.; {Kaspi}, V.M.; {Kaplan}, D.L.
\newblock {Twenty-One Millisecond Pulsars in Terzan 5 Using the Green Bank
  Telescope}.
\newblock {\em Science} {\bf 2005}, {\em 307},~892--896,
\newblock
  doi:{\changeurlcolor{black}\href{https://doi.org/10.1126/science.1108632}{\detokenize{10.1126/science.1108632}}}.

\bibitem[{Hessels} \em{et~al.}(2006){Hessels}, {Ransom}, {Stairs}, {Freire},
  {Kaspi}, and {Camilo}]{2006Sci...311.1901H}
{Hessels}, J.W.T.; {Ransom}, S.M.; {Stairs}, I.H.; {Freire}, P.C.C.; {Kaspi},
  V.M.; {Camilo}, F.
\newblock {A Radio Pulsar Spinning at 716 Hz}.
\newblock {\em Science} {\bf 2006}, {\em 311},~1901--1904,
\newblock
  doi:{\changeurlcolor{black}\href{https://doi.org/10.1126/science.1123430}{\detokenize{10.1126/science.1123430}}}.

\bibitem[{Bogdanov} \em{et~al.}(2011){Bogdanov}, {van den Berg}, {Servillat},
  {Heinke}, {Grindlay}, {Stairs}, {Ransom}, {Freire}, {B{\'e}gin}, and
  {Becker}]{2011ApJ...730...81B}
{Bogdanov}, S.; {van den Berg}, M.; {Servillat}, M.; {Heinke}, C.O.;
  {Grindlay}, J.E.; {Stairs}, I.H.; {Ransom}, S.M.; {Freire},~P.C.C.;
  {B{\'e}gin}, S.; {Becker}, W.
\newblock {Chandra X-ray Observations of 12 Millisecond Pulsars in the Globular
  Cluster M28}.
\newblock {\em  Astrophys. J.} {\bf 2011}, {\em 730},~81,
\newblock
  doi:{\changeurlcolor{black}\href{https://doi.org/10.1088/0004-637X/730/2/81}{\detokenize{10.1088/0004-637X/730/2/81}}}.

\bibitem[{Pallanca} \em{et~al.}(2010){Pallanca}, {Dalessandro}, {Ferraro},
  {Lanzoni}, {Rood}, {Possenti}, {D'Amico}, {Freire}, {Stairs}, {Ransom}, and
  {B{\'e}gin}]{2010ApJ...725.1165P}
{Pallanca}, C.; {Dalessandro}, E.; {Ferraro}, F.R.; {Lanzoni}, B.; {Rood},
  R.T.; {Possenti}, A.; {D'Amico}, N.; {Freire}, P.C.; {Stairs}, I.; {Ransom},
  S.M.; {B{\'e}gin}, S.
\newblock {The Optical Companion to the Binary Millisecond Pulsar J1824-2452H
  in the Globular Cluster M28}.
\newblock {\em  Astrophys. J.} {\bf 2010}, {\em 725},~1165--1169,
\newblock
  doi:{\changeurlcolor{black}\href{https://doi.org/10.1088/0004-637X/725/1/1165}{\detokenize{10.1088/0004-637X/725/1/1165}}}.

\bibitem[{Pallanca} \em{et~al.}(2013){Pallanca}, {Dalessandro}, {Ferraro},
  {Lanzoni}, and {Beccari}]{2013ApJ...773..122P}
{Pallanca}, C.; {Dalessandro}, E.; {Ferraro}, F.R.; {Lanzoni}, B.; {Beccari},
  G.
\newblock {The Optical Counterpart to the X-ray Transient IGR J1824-24525 in
  the Globular Cluster M28}.
\newblock {\em  Astrophys. J.} {\bf 2013}, {\em 773},~122,
\newblock
  doi:{\changeurlcolor{black}\href{https://doi.org/10.1088/0004-637X/773/2/122}{\detokenize{10.1088/0004-637X/773/2/122}}}.

\bibitem[]{papitto2013}
Papitto, A.; Ferrigno, C.; Bozzo, E.; Rea, N.; Pavan, L.; Burderi, L.; Burgay, M.; Campana, S.; di Salvo, T.; Falanga, M.; et al. Swings between rotation and accretion power in a binary millisecond pulsar. {\em Nature} {\bf 2013}, {\em 501}, 517.

\bibitem[{Lynch} \em{et~al.}(2011){Lynch}, {Ransom}, {Freire}, and
  {Stairs}]{2011ApJ...734...89L}
{Lynch}, R.S.; {Ransom}, S.M.; {Freire}, P.C.C.; {Stairs}, I.H.
\newblock {Six New Recycled Globular Cluster Pulsars Discovered with the Green
  Bank Telescope}.
\newblock {\em  Astrophys. J.} {\bf 2011}, {\em 734},~89,
\newblock
  doi:{\changeurlcolor{black}\href{https://doi.org/10.1088/0004-637X/734/2/89}{\detokenize{10.1088/0004-637X/734/2/89}}}.

\bibitem[{Freire} \em{et~al.}(2005){Freire}, {Hessels}, {Nice}, {Ransom},
  {Lorimer}, and {Stairs}]{2005ApJ...621..959F}
{Freire}, P.C.C.; {Hessels}, J.W.T.; {Nice}, D.J.; {Ransom}, S.M.; {Lorimer},
  D.R.; {Stairs}, I.H.
\newblock {The Millisecond Pulsars in NGC 6760}.
\newblock {\em  Astrophys. J.} {\bf 2005}, {\em 621},~959--965,
\newblock
  doi:{\changeurlcolor{black}\href{https://doi.org/10.1086/427748}{\detokenize{10.1086/427748}}}.

\bibitem[{Cadelano} \em{et~al.}(2015){Cadelano}, {Pallanca}, {Ferraro},
  {Stairs}, {Ransom}, {Dalessandro}, {Lanzoni}, {Hessels}, and
  {Freire}]{2015ApJ...807...91C}
{Cadelano}, M.; {Pallanca}, C.; {Ferraro}, F.R.; {Stairs}, I.; {Ransom}, S.M.;
  {Dalessandro}, E.; {Lanzoni}, B.; {Hessels}, J.W.T.; {Freire}, P.C.C.
\newblock {Radio Timing and Optical Photometry of the Black Widow System PSR
  J1953+1846A in the Globular Cluster M71}.
\newblock {\em  Astrophys. J.} {\bf 2015}, {\em 807},~91,
\newblock
  doi:{\changeurlcolor{black}\href{https://doi.org/10.1088/0004-637X/807/1/91}{\detokenize{10.1088/0004-637X/807/1/91}}}.

\bibitem[{Ransom} \em{et~al.}(2004){Ransom}, {Stairs}, {Backer}, {Greenhill},
  {Bassa}, {Hessels}, and {Kaspi}]{2004ApJ...604..328R}
{Ransom}, S.M.; {Stairs}, I.H.; {Backer}, D.C.; {Greenhill}, L.J.; {Bassa},
  C.G.; {Hessels}, J.W.T.; {Kaspi}, V.M.
\newblock {Green Bank Telescope Discovery of Two Binary Millisecond Pulsars in
  the Globular Cluster M30}.
\newblock {\em  Astrophys. J.} {\bf 2004}, {\em 604},~328--338,
\newblock
  doi:{\changeurlcolor{black}\href{https://doi.org/10.1086/381730}{\detokenize{10.1086/381730}}}.




\bibitem[]{hui2014}
Hui, C.Y. Spider Invasion Across the Galaxy. {\em J. Astron. Space Sci.} {\bf 2014}, {\em 31}, 101--120.
\bibitem[]{vv1988}
van den Heuvel, E.P.J.; van Paradijs, J. Fate of the companion stars of ultra-rapid pulsars. {\em Nature} {\bf 1988}, {\em 334}, 227--228.
\bibitem[]{kluzniak1988}
Kluzniak, W.; Ruderman, M.; Shaham, J.; Tavani, M. Nature and evolution of the eclipsing millisecond binary pulsar PSR1957+20. {\em Nature}, {\bf 1988}, {\em 334}, 225--227.
\bibitem[]{ruderman1989a}
Ruderman, M.; Shaham, J.; Tavani, M. Accretion turnoff and rapid evaporation of very light secondaries in low-mass X-ray binaries. {\em \apj} {\bf 1989}, {\em 336}, 507--518.
\bibitem[]{ruderman1989b}
Ruderman, M.; Shaham, J.; Tavani, M.; Eichler, D. Late evolution of very low mass X-ray binaries sustained by radiation from their primaries. {\em \apj} {\bf 1989}, {\em 343}, 292--312.
\bibitem[]{stappers2003}
Stappers, B.W.; Gaensler, B.M.; Kaspi, V.M.; van der Klis, M.; Lewin, W.H.G. An X-ray Nebula Associated with the Millisecond Pulsar B1957+20. {\em Science} {\bf 2003}, {\em 299}, 1372--1374.
\bibitem[]{cheng2006}
Cheng, K.S.; Taam, R.E.; Wang, W. Pulsar Wind Nebulae and the Non-thermal X-ray Emission of Millisecond Pulsars. {\em \apj} {\bf 2006}, {\em 641}, 427--437.
\bibitem[]{kargaltsev2013}
Kargaltsev, O.; Rangelov, B.; Pavlov, G.G. Gamma-ray and X-ray Properties of Pulsar Wind Nebulae and Unidentified Galactic TeV Sources.
In {\em The Universe Evolution. Astrophysical and Nuclear Aspects}; Nova Science Publishers, Inc.: New York, NY, USA,  {2013}
\bibitem[]{hui2006}
Hui, C.Y.; Becker, W. Searches for diffuse X-ray emission around millisecond pulsars:
An X-ray nebula associated with PSR J2124-3358. {\em Astron. Astrophys.} {\bf 2006}, {\em 448}, L13--L17.
\bibitem[]{lee2018a}
Lee, J.; Hui, C. Y.; Takata, J.; Lin, L.C.C. Discovery of an X-ray nebula in the field of millisecond pulsar
PSRJ1911--1114. {\em Astron. Astrophys.} {\bf 2018}, {\em 620}, L14.
\bibitem[]{homer2006}
Homer, L.; Szkody, P.; Chen, B.; Henden, A.; Schmidt, G.; Anderson, S.F.; Silvestri, N.M.; Brinkmann, J.
 XMM-Newton and Optical Follow-up Observations of SDSS J093249.57+472523.0 and SDSS J102347.67+003841.2. {\em Astron. J.} {\bf 2006}, {\em 131}, 562--570.
\bibitem[]{wang2009} Wang, Z.; Archibald, A.M.; Thorstensen, J.R.; Kaspi, V.M.; Lorimer, D.R.; Stairs, I.; Ransom, S.M. SDSS J102347.6+003841: A Millisecond Radio Pulsar Binary That Had a Hot Disk During 2000-2001. {\em \apj} {\bf 2009}, {\em 703}, 2017--2023.
\bibitem[]{tam2010}
Tam, P.H.T.; Hui, C.Y.; Huang, R.H.H.; Kong, A.K.H.; Takata, J.; Lin, L.C.C.; Yang, Y.J.; Cheng, K.S.; Taam, R.E. Evidence for gamma-ray emission from the low-mass x-ray binary
system first J102347.6+003841. {\em \apjl} {\bf 2010}, {\em 724}, L207--L211.
\bibitem[]{stappers2013} Stappers, B.W.;  Archibald, A.; Bassa, C.; Hessels, J.; Janssen, G.; Kaspi, V.; Lyne, A.; Patruno, A.; Hill, A.B.
State-change in the "transition" binary millisecond pulsar J1023+0038. {\em Astron. Telegr.} {\bf 2013}, {\emph 5513}. 
\bibitem[]{patruno2014} Patruno, A.; Archibald, A.M.; Hessels, J.W.T.; Bogdanov, S.; Stappers, B.W.; Bassa, C.G.; Janssen, G.H.; Kaspi, V.M.; Tendulkar, S.; Lyne, A.G. A New Accretion Disk around the Missing Link Binary System PSR J1023+0038. {\em \apj} {\bf 2014}, {\em 781}, L3.
\bibitem[]{halpern2013}
Halpern, J.P.; Gaidos, E.; Sheffield, A.; Price-Whelan, A.M.; Bogdanov, S. Optical Observations of the Binary MSP J1023+0038 in a New Accreting State. {\em Astron. Telegr.} {\bf 2013}, {\em 5514}, 1.
\bibitem[]{bogdanov2011}
Bogdanov, S.; Archibald, A.M.; Hessels, J.W.T.; Kaspi, V.M.; Lorimer, D.; McLaughlin, M.A.; Ransom, S.M.; Stairs, I.H. A Chandra X-ray Observation of the Binary Millisecond Pulsar PSR J1023+0038. {\em \apj} {\bf 2011}, {\em 742}, 97.
\bibitem[]{deMartino2010}
de Martino D.; Falanga, M.; Bonnet-Bidaud, J.-M.; Belloni, T.; Mouchet, M.; Masetti, N.; Andruchow, I.; Cellone, S.A.; Mukai, K.; Matt, G. The intriguing nature of the high-energy gamma ray source XSS J12270-4859. {\em Astron.~Astrophys.} {\bf 2010}, {\em 515}, 25.
\bibitem[]{bassa2013}
Bassa, C.G.; Patruno, A.; Hessels, J.W.T.; Archibald, A.M.; Mahony, E.K.; Monard, B.; Keane, E.F.; Bogdanov, S.; Stappers, B.W.; Janssen, G.H.; et al. A possible state transition in the low-mass X-ray binary XSS J12270-4859. {\em Astron. Telegr.} {\bf 2013},~\emph {5647}. 
\bibitem[]{tam2013}
Tam, P.; Kong, A.; Li, K. Fermi/LAT and Swift/XRT observations of XSS J12270-4859/2FGL J1227.7-4853. {\em Astron. Telegr.} {\bf 2013}, \emph {5652}.
\bibitem[]{roy2014}
Roy, J.; Bhattacharyya, B.; Ray, P. GMRT discovery of a 1.69 ms radio pulsar associated with XSS J12270-4859 {\em Astron. Telegr.} {\bf 2014}, \emph {5890}.
\bibitem[]{hui2015b}
Hui, C.Y.; Hu, C.P.; Park, S.M.; Takata, J.; Li, K.L.; Tam, P.H.T.; Lin, L.C.C.; Kong, A.K.H.; Cheng, K.S.; Kim, C. Exploring the Intrabinary Shock from the Redback Millisecond Pulsar PSR J2129-0429. {\em \apj} {\bf 2015}, {\em 801}, L27.
\bibitem[]{possenti2002} Possenti, A.; Cerutti, R.; Colpi, M.; Mereghetti, S. Re-examining the X-ray versus spin-down luminosity correlation of rotation powered pulsars. {\em Astron. Astrophys.} {\bf 2002}, {\em 387}, 993.
\bibitem[]{hui2010} Hui, C.Y.; Cheng, K.S.; Taam, R.E. Dynamical Formation of Millisecond Pulsars in Globular Clusters. {\em \apj} {\bf 2010}, {\em 714}, 1149--1154.
\bibitem[]{parkinson2016}
Saz Parkinson, P.M.; Xu, H.; Yu, P.L.H.; Salvetti, D.; Marelli, M.; Falcone, A.D. Classification and ranking of Fermi LAT gamma-ray sources from the 3FGL catalog using machine learning techniques.
{\em \apj} {\bf 2016}, {\em 820}, 8.
\bibitem[]{leung2017}
Leung, A.P.; Tong, Y.; Li, R.; Luo, S.; Hui, C.Y. A Novel Framework for Gamma-ray Source Classification using Automatic Feature Selection.
{\em Proc. Sci.} {\bf 2017}, {\em 312}, 133.
\bibitem[{{Li} {et~al.}(2016){Li}, {Kong}, {Hou}, {Mao}, {Strader}, {Chomiuk},
  \& {Tremou}}]{2016ApJ...833..143L}
Li, K.-L.; Kong, A.K.H.; Hou, X.; Mao, J.; Strader, J.; Chomiuk, L.; Tremou, E. Discovery of a Redback Millisecond Pulsar Candidate: 3FGL~J0212.1+5320. {\em \apj} {\bf 2016}, {\em 833}, 143.
\bibitem[{{Linares} {et~al.}(2017){Linares}, {Miles-P{\'a}ez},
  {Rodr{\'{\i}}guez-Gil}, {Shahbaz}, {Casares}, {Fari{\~n}a}, \&
  {Karjalainen}}]{2017MNRAS.465.4602L}
{Linares}, M.; {Miles-P{\'a}ez}, P.; {Rodr{\'{\i}}guez-Gil}, P.; Shahbaz, T.; Casares, J.; Fariña, C.; Karjalainen, R.
A millisecond pulsar candidate in a 21-h orbit: 3FGL~J0212.1+5320.
{\em Mon. Not. R.  Astron. Soc.} {\bf 2017}, {\em 465}, 4602.
\bibitem[{{Strader} {et~al.}(2016){Strader}, {Li}, {Chomiuk}, {Heinke},
  {Udalski}, {Peacock}, {Shishkovsky}, \& {Tremou}}]{2016ApJ...831...89S}
Strader, J.; Li, K.-L.; Chomiuk, L.; Heinke, C.O.; Udalski, A.; Peacock, M.; Shishkovsky, L.; Tremou, E.
A New gamma-Ray Loud Eclipsing Low-mass X-ray Binary.
{\em \apj} {\bf 2016}, {\em 831}, 89.
\bibitem[{{Strader} {et~al.}(2014){Strader}, {Chomiuk}, {Sonbas}, {Sokolovsky},
  {Sand}, {Moskvitin}, \& {Cheung}}]{2014ApJ...788L..27S}
Strader, J.; Chomiuk, L.; Sonbas, E.; Sokolovsky, K.; Sand, D.J.; Moskvitin, A.S.; Cheung, C.C.
1FGL J0523.5-2529: A New Probable Gamma-Ray Pulsar Binary.
{\em \apj} {\bf 2014}, {\em 788}, L27.

\bibitem[{Xing} \em{et~al.}(2014){Xing}, {Wang}, and {Ng}]{2014ApJ...795...88X}
{Xing}, Y.; {Wang}, Z.; {Ng}, C.Y.
\newblock {Fermi Variability Study of the Candidate Pulsar Binary 2FGL
  J0523.3-2530}.
\newblock {\em  Astrophys. J.} {\bf 2014}, {\em 795},~88,
\newblock
  doi:{\changeurlcolor{black}\href{https://doi.org/10.1088/0004-637X/795/1/88}{\detokenize{10.1088/0004-637X/795/1/88}}}.

\bibitem[{{Salvetti} {et~al.}(2017){Salvetti}, {Mignani}, {De Luca}, {Marelli},
  {Pallanca}, {Breeveld}, {H{\"u}semann}, {Belfiore}, {Becker}, \&
  {Greiner}}]{2017MNRAS.470..466S}
Salvetti, D.; Mignani, R.P.; De Luca, A.; Marelli, M.; Pallanca, C.; Breeveld, A.A.; H\"{u}semann, P.; Belfiore, A.; Becker, W.; Greiner, J.
A multiwavelength investigation of candidate millisecond pulsars in unassociated gamma-ray sources.
{\em  Mon. Not. R. Astron. Soc.} {\bf 2017}, {\em 470}, 466
\bibitem[{{Halpern} {et~al.}(2017){Halpern}, {Strader}, \&
  {Li}}]{2017ApJ...844..150H}
{Halpern}, J.P.; {Strader}, J.; {Li}, M.
A Likely Redback Millisecond Pulsar Counterpart of 3FGL J0838.8-2829.
{\em \apj} {\bf 2017}, {\em 844}, 150.
\bibitem[{{Swihart} {et~al.}(2017){Swihart}, {Strader}, {Johnson}, {Cheung},
  {Sand}, {Chomiuk}, {Wasserman}, {Larsen}, {Brodie}, {Simonian}, {Tremou},
  {Shishkovsky}, {Reichart}, \& {Haislip}}]{2017ApJ...851...31S}
Swihart, S.J.; Strader, J.; Johnson, T.J.; Cheung, C.C.; Sand, D.; Chomiuk, L.; Wasserman, A.; Larsen, S.; Brodie, J.P.; Simonian, G.V.; et~al.
2FGL J0846.0+2820: A New Neutron Star Binary with a Giant Secondary and Variable $\gamma$-Ray Emission.
{\em \apj} {\bf 2017}, {\em 851}, 31.
\bibitem[{{Li} {et~al.}(2018){Li}, {Hou}, {Strader}, {Takata}, {Kong},
  {Chomiuk}, {Swihart}, {Hui}, \& {Cheng}}]{2018ApJ...863..194L}
Li, K.-L.; Hou, X.; Strader, J.; Takata, J.; Kong, A.K.H.; Chomiuk, L.; Swihart, S.J.; Hui, C.Y.; Cheng, K.S.
Multiwavelength Observations of a New Redback Millisecond Pulsar Candidate: 3FGL J0954.8-3948.
{\em \apj} {\bf 2018}, {\em 863}, 194. 
\bibitem[{{Coti Zelati} {et~al.}(2019){Coti Zelati}, {Papitto}, {de Martino},
  {Buckley}, {Odendaal}, {Li}, {Russell}, {Torres}, {Mazzola}, \&
  {Bozzo}}]{2019AA...622A.211C}
Coti Zelati, F.; Papitto, A.; de Martino, D.; Buckley, D.A.H.; Odendaal, A.; Li, J.; Russell, T.D.; Torres, D.F.; Mazzola, S.M.; Bozzo, E.; et~al.
Prolonged sub-luminous state of the new transitional pulsar candidate CXOU J110926.4-650224.
{\em Astron. Astrophys.} {\bf 2019}, {\em 622}, A211.
\bibitem[{{Bogdanov} \& {Halpern}(2015)}]{2015ApJ...803L..27B}
{Bogdanov}, S.; {Halpern}, J.P.
Identification of the High-energy Gamma-Ray Source 3FGL J1544.6-1125 as a Transitional Millisecond Pulsar Binary in an Accreting State.
{\em \apj} {\bf 2015}, {\em 803}, L27.
\bibitem[{{Britt} {et~al.}(2017){Britt}, {Strader}, {Chomiuk}, {Tremou},
  {Peacock}, {Halpern}, \& {Salinas}}]{2017ApJ...849...21B}
{Britt}, C.T.; {Strader}, J.; {Chomiuk}, L.; Tremou, E.; Peacock, M.; Halpern, J.; Salinas, R. Orbital Dynamics of Candidate Transitional Millisecond Pulsar 3FGL J1544.6-1125: An Unusually Face-on System.
{\em \apj} {\bf 2017}, {\em 849}, 21.
\bibitem[{{Romani} {et~al.}(2014){Romani}, {Filippenko}, \&
  {Cenko}}]{2014ApJ...793L..20R}
{Romani}, R.W.; {Filippenko}, A.V.; {Cenko}, S.B.
2FGL J1653.6-0159: A New Low in Evaporating Pulsar Binary Period.
{\em \apj} {\bf 2014}, {\em 793}, L20.
\bibitem[{{Kong} {et~al.}(2014){Kong}, {Jin}, {Yen}, {Hu}, {Hui}, {Tam},
  {Takata}, {Lin}, {Cheng}, \& {Park}}]{2014ApJ...794L..22K}
Kong, A.K.H.; Jin, R.; Yen, T.-C.; Hu, C.-P.; Hui, C.Y.; Tam, P.H.T.; Takata, J.; Lin, L.C.C.; Cheng, K.S.; Park, S.M.; et~al.
Discovery of an Ultracompact Gamma-Ray Millisecond Pulsar Binary Candidate.
{\em \apj} {\bf 2014}, {\em 794}, L22.
\bibitem[{{Romani}(2015)}]{2015ApJ...812L..24R}
{Romani}, R.W. A Likely Millisecond Pulsar Binary Counterpart for Fermi Source 2FGL J2039.6-5620.
{\em \apj} {\bf 2015}, {\em 812}, L24.
\bibitem[{{Salvetti} {et~al.}(2015){Salvetti}, {Mignani}, {De Luca}, {Delvaux},
  {Pallanca}, {Belfiore}, {Marelli}, {Breeveld}, {Greiner}, \&
  {Becker}}]{2015ApJ...814...88S}
Salvetti, D.; Mignani, R.P.; De Luca, A.; Delvaux, C.; Pallanca, C.; Belfiore, A.; Marelli, M.; Breeveld, A.A.; Greiner, J.; Becker, W.; et~al.
Multi-wavelength Observations of 3FGL J2039.6-5618: A~Candidate Redback Millisecond Pulsar.
{\em \apj} {\bf 2015}, {\em 814}, 88.

\bibitem[{Ng} \em{et~al.}(2018){Ng}, {Takata}, {Strader}, {Li}, and
  {Cheng}]{2018ApJ...867...90N}
{Ng}, C.W.; {Takata}, J.; {Strader}, J.; {Li}, K.L.; {Cheng}, K.S.
\newblock {Evidence on the Orbital Modulated Gamma-Ray Emissions from the
  Redback Candidate 3FGL J2039.6-5618}.
\newblock {\em  Astrophys. J.} {\bf 2018}, {\em 867},~90,
\newblock
  doi:{\changeurlcolor{black}\href{https://doi.org/10.3847/1538-4357/aae308}{\detokenize{10.3847/1538-4357/aae308}}}.

\bibitem[]{arons1993}
Arons, J.; Tavani, M.
High-energy emission from the eclipsing millisecond pulsar PSR 1957+20.
{\em \apj} {\bf 1993}, {\em 403}, 249.
\bibitem[]{bednarek2013}
Bednarek, W.; Sitarek, J.
High-energy emission from the nebula around the Black Widow binary system containing millisecond pulsar B1957+20.
{\em Astron. Astrophys.} {\bf 2013}, {\em 550}, A39.
\bibitem[]{ahnen2017}
Ahnen, M.L.; Ansoldi, S.; Antonelli, L.A.; Arcaro, C.; Babi\'{c}, A.; Banerjee, B.; Bangale, P.; Barres de Almeida, U.; Barrio, J.A.; Becerra Gonz\`{a}lez, J.; et al.
Observation of the black widow B1957+20 millisecond pulsar binary system with the MAGIC telescopes.
{\em  Mon. Not. R. Astron. Soc.} {\bf 2017}, {\em 470}, 4608--4617.

\end{thebibliography}



\end{document}